\documentclass[12pt]{iopart}
\usepackage{cite}
\usepackage{pifont}
\usepackage{amssymb}
\usepackage{iopams}      
\usepackage{graphicx}    
\usepackage{times,txfonts}
\usepackage{mathrsfs}
\usepackage{color}
\usepackage[dvipsnames]{xcolor} 
\usepackage[colorlinks=true,linkcolor=blue,urlcolor=blue,citecolor=blue,pdfusetitle]{hyperref}
\let\iint\relax

\usepackage{amsmath} 
\newcommand{\wht}[1]{\widehat{#1}}
\newcommand{\wit}[1]{\widetilde{#1}}

\newcommand{\bigqm}{\textbf{\large ?}}
\newcommand{\cmark}{\ding{51}} 
\newcommand{\xmark}{\ding{55}}
\newcommand{\omark}{\ding{108}}
\newcommand{\II}{{\rm i}}
\newcommand{\demi}{\frac{1}{2}}
\newcommand{\D}{{\rm d}}
\newcommand{\vep}{\varepsilon}
\newcommand{\bigcdot}
{\mathbin{\vcenter{\hbox{\scalebox{1.5}{$\cdot$}}}}}
\newcommand{\vek}[1]{\boldsymbol{#1}}
\renewcommand{\Vec}[1]{\vek{#1}}

\begin{document}

\title[]{Limitations of the Markovian approximation in the harmonic oscillator
}

\author{Michele Coppola$^{1,2,\star}$, Zoubair Daouma$^{1,3}$ and Malte Henkel$^{1,4}$}
\vspace{10pt}
\address{$^1$Laboratoire de Physique et Chimie Th\'eoriques (CNRS UMR 7019),  Universit\'e de Lorraine Nancy, B.P. 70239, F -- 54506 Vand{\oe}uvre-l\`es-Nancy Cedex, France}
\vspace{10pt}
\address{$^2$Jo\v{z}ef Stefan Institute, SI -- 1000 Ljubljana, Slovenia}
\vspace{10pt}
\address{$^3$Laboratoire de Physique des Lasers, Atomes et Molécules (CNRS UMR 8523), Université de Lille, F -- 59000 Lille, France}
\vspace{10pt}
\address{$^4$Centro de F\'{i}sica Te\'{o}rica e Computacional, Universidade de Lisboa, Campo Grande, P--1749-016 Lisboa, Portugal}
\vspace{10pt}
\ead{\hspace{0.5cm}michele.coppola@ijs.si \hspace{0.3cm} malte.henkel@univ-lorraine.fr}
\vspace{10pt}

\begin{abstract}
The quantum fluctuation-dissipation theorem is a central ingredient in the construction of quantum dynamics of Brownian motion which necessarily is {\em non}-Markovian. 
Yet, often Markovian approximations to quantum dynamics are studied in the literature.
In this work, we investigate the limitations of the Markovian approximation within two paradigmatic models describing a single damped harmonic oscillator. 
These models are governed by distinct quantum Langevin equations, although both are constructed to satisfy the same set of phenomenological criteria: 
the canonical commutation relations between position and momentum, 
the Kubo response relation, the virial theorem, and the equilibrium quantum variance. 
The limitations of the Markovian approximation are underscored by the classical limit, violations of the Ehrenfest theorem, 
the breakdown of complete and simple positivity in the reconstructed master equations, and anomalies in thermalisation behaviour. 
Further phenomenological differences between the two models are illustrated through their quantum relaxation dynamics and phase diagrams, 
derived from their reinterpretation as mean-field approximations of a many-body interacting magnet. 
Our analysis explicitly reveals intrinsic inconsistencies introduced by the Markovian approximation, 
emphasising the need for {\em non}-Markovian frameworks for a consistent description of open quantum dynamics.
\end{abstract}
\vspace{5pt}
\address{$\star$ Corresponding author}

\section{Introduction}

\begin{table}
\caption{Physical requirements for friction and cavity models are indicated as fulfilled (\cmark), violated only under specific conditions (\omark), or violated (\xmark). The question mark ($\boldsymbol{?}$) is meant to indicate that no known classical system emerges for the cavity model.}
\label{fig:tab}
\begin{indented}
\item[]\begin{tabular}{@{}p{4.5cm}p{3.8cm}p{3.8cm}}
\br
\textbf{Requirement}             & \textbf{Cavity Model} & \textbf{Friction Model} \\
\mr
Classical limit                  & \qquad$\bigqm$ & Brownian motion \\
Ehrenfest theorem                & \qquad\xmark   &\;\;\;\qquad \cmark \\
Spatial translation-invariance   & \qquad\xmark   &\;\;\;\qquad \cmark \\
Complete positivity              & \qquad\cmark   & \;\;\;\qquad\xmark \\
Simple positivity                & \qquad\cmark   & \;\;\;\qquad\omark \\
Thermalisation                   & \qquad\omark   & \;\;\;\qquad\cmark \\
\br
\end{tabular}
\end{indented}
\end{table}

The description of the non-equilibrium dynamics of open quantum systems continues to present many challenges 
\cite{cohen2001,breuer2002theory,10.1007/978-3-540-44835-8_7,gardiner2004quantum,schaller2014open,tauber2014critical,Caldeira2014,breuer2016colloquium,de2017dynamics,alicki2018,Weiss2022}. 
In spite of much progress, many details of the modelling of the interaction between the degrees of freedom of the system and the bath(s) are far from obvious. 
In the celebrated system-interaction-bath framework, exact quantum Langevin equations can be derived. {}From these microscopic derivations, 
the {\em quantum fluctuation-dissipation theorem} ({\sc qfdt}) follows, such that the relaxation towards the thermodynamic quantum equilibrium state is guaranteed, 
but the resulting dynamics cannot be Markovian ~\cite{ford1965,ford1988,gardiner2004quantum,hanggi2005fundamental,sieberer2015thermodynamic,ford2017fluctuation,araujo2019axiomatic,Weiss2022}. 
This work aims to deepen our understanding of the limitations of the Markovian hypothesis—which assumes weak system–environment correlations and a clear separation between system timescales 
and bath correlation times—within two paradigmatic models of the damped harmonic oscillator.

Although the harmonic oscillator has served as a paradigm for quantum dynamics for a long time, e.g. 
Refs.~\cite{lindblad1976brownian,lindblad1976generators,gorini1976,CaldeiraLegget1983,kohen1997,cohen2001,bedeaux2001mesoscopic,bedeaux2002non,carmichael1999statistical,breuer2002theory,gardiner2004quantum,2020JSMTE..02.3106D,Rama2009,Weiss2022,alicki2023}, surprises might still be found. 
Notably, it apparently passed unnoticed that the two most common formulations of the quantum damped harmonic oscillator are not equivalent, 
since their two-point functions show different dynamics—despite of both being derived from the Redfield theory~\cite{kohen1997}. One, for clarity referred to
as the {\em cavity model}, stems from quantum optics~\cite{carmichael1999statistical} and is described by a Lindblad equation, while the other one, referred to as the {\em friction model}, 
is derived from the Langevin formalism of a single damped particle~\cite{CaldeiraLegget1983,bedeaux2001mesoscopic,bedeaux2002non}. 
In this work, we show that the Markovian approximation inevitably produces inconsistencies in both models, 
either by violating fundamental principles of quantum mechanics or by contradicting expected behaviour of observables, thus 
highlighting the necessity to go beyond this approximation. 
A summary of these inconsistencies is provided in Tab.~\ref{fig:tab}.\footnote{Any attempt of interpolation between the two models would inherit the inconsistencies of both.} 
First, while the classical limit $\hbar\to 0$ of the friction model converges to an ohmically damped particle 
in a harmonic potential subjected to white noise—consistent with Ehrenfest's theorem—no analogous known classical counterpart arises for the cavity model, for which Ehrenfest's theorem does not hold. 
Second, while the cavity model is described by a well-established Lindblad equation—a {\em completely positive} 
({\sc cp}), time-local form~\cite{breuer2002theory,lindblad1976generators,gorini1976,alicki2007quantum,doi:10.1080/01969720151033490,alicki1995comment,bondar2016,alicki2018,dilley2021,sargolzahi2022linearity,alicki2023} 
(see Ref.~\cite{alicki2017solarcell} for applications to solar and fuel cells)—the friction model’s dynamical propagator is not {\sc cp} \cite{CaldeiraLegget1983,kohen1997,Rama2009}.  
Third, although the friction model preserves both translation-invariance\footnote{We follow Ref.~\cite{kohen1997}, where `translation-invariance' means that the dissipative terms in the Langevin equations should be position-independent.}  and quantum-classical correspondence, 
its dynamical propagator can potentially lead to a breakdown of positivity for sufficiently squeezed initial states~\cite{talkner1986}. 
Fourth, thermalisation is consistently guaranteed in the friction model—with positivity nonetheless always being restored in the long-time limit—whereas the Gibbs steady state of the cavity model 
exhibits re-entrant damping-dependent contributions that prevent thermalisation in the presence of magnetic fields. 
These are some of the consequences of the physically inadmissible Markov approximation, see again Tab.~\ref{fig:tab}. 
These difficulties can only be avoided by using a full quantum dynamics
which by nature is {\em non}-Markovian \cite{hanggi2005fundamental,sieberer2015thermodynamic,araujo2019axiomatic}.

More specifically, in this work we shall proceed as follows. 
First, we employ the Langevin formalism to define the cavity and friction models, both special cases of a general damped harmonic oscillator studied long ago via quantum master 
equations~\cite{lindblad1976brownian,CaldeiraLegget1983,kohen1997}, where the noise terms are chosen to ensure that certain observable properties of the stationary state and the system dynamics are 
preserved~\cite{Nonequilibrium-Zwanzig,2020JSMTE..02.3106D,Weiss2022}: (A) the canonical equal-time commutator, (B) the Kubo formula of linear response theory, 
(C) the virial theorem and (D) the variance at equilibrium for quantum harmonic oscillators, see Sec.~\ref{sec:mod}. 
This phenomenological approach circumvents any explicit discussion of initial system-bath correlations while ensuring that both models satisfy the same physical 
properties.\footnote{The admissibility of the tensor product form for initial states, which is a sufficient condition for generating {\sc cp} quantum mechanical maps~\cite{breuer2002theory}, 
has been questioned~\cite{pechukas1994reduced, pechukas1995pechukas, romero2004dynamics, shaji2005s, carteret2008dynamics, rodriguez2008completely, dominy2016beyond}, 
as well as, more broadly, the necessity of complete positivity as a dynamical postulate. Opinions range from viewing complete positivity as a fundamental 
requirement~\cite{lindblad1976brownian,lindblad1976generators,alicki1995comment,alicki2007quantum,sargolzahi2022linearity,manzano2020,alicki2023}, 
to arguing that it is unnecessary in the presence of initial system-bath correlations~\cite{pechukas1994reduced, pechukas1995pechukas, shaji2005s,Colla2022}, 
yet being still valid for classical system-bath correlations~\cite{rodriguez2008completely, carteret2008dynamics} 
but incompatible with Ehrenfest's theorem~\cite{bondar2016} and either with spatial translation-invariance or else with 
relaxation towards equilibrium~\cite{lindblad1976brownian,kohen1997,tupkary2022}. Others recognize the existence of infinitely many {\sc cp} dynamics, but of measure zero in the larger ensemble of 
dynamics~\cite{jagadish2020}, or even that any dynamics can be made {\sc cp}~\cite{dilley2021}. 
Alternatively, the whole question has been dismissed and one calls for a conceptually totally new approach~\cite{Schmid2019}. 
This intense and on-going debate has even raised the interest of philosophers \cite{cuffaro2013}.}
Second, we seek the effective master equations that replicate the equations of motion for the quadratic observables obtained from the Langevin formalism. 
If the time-local generator governing the oscillator’s dynamics is {\sc cp}, as in the cavity model, a standard argument~\cite{lindblad1976generators,gorini1976} 
ensures that the oscillator’s density matrix remains positive throughout the quantum evolution. In contrast, when the dynamical propagator fails to 
preserve even basic positivity—as observed in the friction model—this signals that the Markovian approximation is not consistent across all classes 
of initial system density matrices. A comprehensive analysis of the underlying physical mechanisms responsible for this positivity breakdown lies beyond the scope of the present study. 
Instead, third, we focus on highlighting the limitations of the Markovian approximation through the phase-space formulation of quantum mechanics, 
which provides a unified framework to characterise the relaxation dynamics and the asymptotic convergence of both models towards a physically admissible, positive equilibrium density matrix.
The simultaneous use of all three formulations is our tool to arrive at the physical insight of the models' inconsistencies as listed in Tab.~\ref{fig:tab}.

Specifically, our programme will be as follows. In Sec.~\ref{sec:mod}, the friction model is defined as the Bedeaux-Mazur formulation \cite{bedeaux2001mesoscopic,bedeaux2002non}
of the Markovian quantum Langevin equations of the damped harmonic oscillator, while the cavity model is defined as a quantum harmonic oscillator with different deterministic damping terms. 
In both models, the explicit noise correlators are determined by the same phenomenological criteria. 
Therefore, regardless of their microscopic origins, both models must satisfy identical observable phenomenological requirements, 
distinguishing this approach from the abstract, mathematically motivated constructs commonly found in the literature. 
The exact solutions of the Langevin equations for both models are also presented, with the friction model exhibiting an over-damped dynamical regime that is absent in the cavity model. 
In Sec.~\ref{sec:mast}, not least in order to relate our work with a celebrated comparative study of quantum master equations~\cite{kohen1997}, we recall
that the cavity model is equivalent to the standard Lindbladian description of a harmonic oscillator in a cavity ~\cite{carmichael1999statistical,breuer2002theory,gardiner2004quantum} 
and hence manifestly {\sc cp}~\cite{alicki2007quantum,alicki2018,alicki2023}, in contrast to the constructed dissipator of the friction model,
which is equivalent to the Caldeira-Leggett model in the large temperature limit~\cite{CaldeiraLegget1983} and well-known to be not {\sc cp}, for which we shall give a new proof.
In Sec.~\ref{sec:dyn}, the master equations of both models, recast as Fokker-Planck equations for the Wigner functions, are explicitly solved. 
The two models exhibit distinct relaxation times and in an external field, they ultimately converge to different Gibbs states. 
Consequently, while the friction model may violate the positivity of the density matrix at relatively short times, 
positivity is always restored in the long-time limit. However, we demonstrate that thermalisation is consistently achieved only in the friction model, 
while for the cavity model it occurs solely in the absence of a magnetic field. 
In Sec.~\ref{sec:mean}, as a physical illustration, we recast both models as mean-field approximations of a many-body interacting magnet, and analyse
the nature of their respective relaxations and phase diagrams at zero temperature. 
While the friction model relaxes to an equilibrium state independent of the damping parameter $\gamma$, 
the cavity model produces a non-trivial re-entrant phase diagram and its $\gamma$-dependent stationary state cannot reach equilibrium. 
Our conclusions are given in Sec.~\ref{sec:disc}. Several appendices contain technical details and background. 

\section{Quantum Langevin equations}
\label{sec:mod}

Our study concerns a single harmonic oscillator, with Hamiltonian  
\begin{equation} \label{eq:H}
    \hat{H}=\,\frac{g}{2}\hat{p}^2+\frac{\omega^2}{2g}\hat{x}^2 - B \hat{x}\,,
\end{equation}
where $g$ is the quantum coupling, $\omega$ is the angular frequency and $\hat{x},\hat{p}$ 
are the position and momentum operators which obey the commutation relation $\left[\hat{x},\hat{p}\right]=\II\hbar$. In application to quantum magnets, 
where $\hat{x}$ becomes a quantum spin variable $\hat{s}$, $B$ is interpreted as an external magnetic field \cite{araujo2019axiomatic}. 
Position and momentum operators can be expressed through bosonic creation and annihilation operators 
\begin{equation} \label{eq:aa}
    \hat{x}=\sqrt{\frac{\hbar g}{2\omega}\,}\,\bigl(\hat{a}^{\dag}+\hat{a}\bigr)\,,\qquad
    \hat{p}=\II\sqrt{\frac{\hbar\omega}{2g}\,}\,\bigl(\hat{a}^\dagger-\hat{a}\bigr)\,.
\end{equation}
which lead to
\begin{equation}\label{eq:H2}
\hat{H}=\hbar\omega\bigg(\hat{a}^\dagger\hat{a}+\frac{1}{2}\bigg) - B \sqrt{\frac{\hbar g}{2\omega}\,}\,(\hat{a}+\hat{a}^\dagger)\,.
\end{equation}
In this work, we compare the dynamics generated by two different sets of Langevin equations, namely the {\em friction model} with
\begin{subequations} \label{eq:dynBM}
\begin{align}
\partial_t \hat{x} &= \frac{\II}{\hbar} \left[\hat{H},\hat{x}\right] + \hat{\eta}^{(x)}_f(t)\,, \label{eq:dynBMa}\\ 
\partial_t \hat{p} &= \frac{\II}{\hbar} \left[\hat{H},\hat{p}\right] - \gamma \hat{p} + \hat{\eta}^{(p)}_f(t)\,, \label{eq:dynBMb}
\end{align}
\end{subequations}
and the {\em cavity model} of quantum optics with\footnote{We follow the terminology of \cite{kohen1997}: since (\ref{eq:dynCava}) 
does contain the coordinate $\hat{x}$, the cavity model is {\em not} spatially translation-invariant, but the friction model is, see (\ref{eq:dynBMa}).}
\begin{subequations} \label{eq:dynCav} 
\begin{align}
\partial_t \hat{x} &= \frac{\II}{\hbar} \left[\hat{H},\hat{x}\right] - \frac{\gamma}{2} \hat{x} + \hat{\eta}^{(x)}_c(t)\,, \label{eq:dynCava} \\ 
\partial_t \hat{p} &= \frac{\II}{\hbar} \left[\hat{H},\hat{p}\right] - \frac{\gamma}{2} \hat{p} + \hat{\eta}^{(p)}_c(t)\,, \label{eq:dynCavb} 
\end{align}
\end{subequations}
for the Hamiltonian $\hat{H}$ of the harmonic oscillator~\eqref{eq:H}. These equations combine the Heisenberg equations of motion, 
describing the unitary part of the dynamics, with phenomenological damping terms parametrized by the damping constant $\gamma$. 
If one places oneself in the context of a particle (of mass $m=g^{-1}$) subject to a damping force which depends linearly on the velocity, the
choice (\ref{eq:dynBMa}, \ref{eq:dynBMb}) might appear a natural one, whereas the choice (\ref{eq:dynCava}, \ref{eq:dynCavb}) may appear surprising. 
However, as we shall recall in Sec.~\ref{sec:mast}, the dynamics (\ref{eq:dynCava}, \ref{eq:dynCavb}) 
reproduces the same correlators of the celebrate Lindblad equation of an oscillator in a 
cavity~\cite{kohen1997,carmichael1999statistical,breuer2002theory,gardiner2004quantum,alicki2007quantum,alicki2023}. 
Finally, the coupling with the external bath(s) is completed by specifying the stochastic noises $\hat{\eta}^{(x)}_\alpha(t)$, $\hat{\eta}^{(p)}_\alpha(t)$, 
for the friction ($\alpha=f$) and cavity ($\alpha=c$) models, which are operator-valued centred random processes with a joint probability distribution and non-vanishing second moments. 

The form of the noise correlators in the friction model~(\ref{eq:dynBMa}, \ref{eq:dynBMb}) follows the proposal by Bedeaux and Mazur~\cite{bedeaux2001mesoscopic,bedeaux2002non}.  
Physical examples are also known from LRC electric circuits with noise both in potential and in current~\cite{becker2013theorie,araujo2019axiomatic} or
from models of active processes in the inner ear~\cite{dinis2012fluctuation}. The non-vanishing noise correlators are \cite{bedeaux2001mesoscopic,bedeaux2002non}

\addtocounter{equation}{-2} 

\begin{subequations}
\addtocounter{equation}{2}
\begin{align}
\bigl\langle \left[ \hat{\eta}^{(x)}_f(t), \hat{\eta}^{(p)}_f(t') \right] \bigr\rangle &= \II \gamma \hbar\, \delta(t-t')\,, \label{eq:dynBMc} \\
\bigl\langle \left\{ \hat{\eta}^{(p)}_f(t), \hat{\eta}^{(p)}_f(t') \right\} \bigr\rangle &= 
              \frac{2\gamma \hbar\omega}{g}\coth\left(\frac{\hbar \omega}{2T}\right)\, \delta(t-t')\,, \label{eq:dynBMd} 
\end{align}
\end{subequations}
with the notations $\bigl[ \cdot, \cdot\bigr]$ and $\{\cdot,\cdot\bigr\}$ for the commutator and anti-commutator (use units with $k_{\rm B}=1$). 
In the classical limit $\hbar\to 0$, this reduces to a linearly damped harmonic oscillator subject to white noise. 
On the other hand, for the cavity model~(\ref{eq:dynCava}, \ref{eq:dynCavb}) we have by definition
\begin{subequations}
\addtocounter{equation}{2}
\begin{align}
\bigl\langle \left[ \hat{\eta}^{(x)}_c(t), \hat{\eta}^{(p)}_c(t') \right] \bigr\rangle &= \II \gamma \hbar\, \delta(t-t')\,, \label{eq:dynCavc} \\
\bigl\langle \left\{ \hat{\eta}^{(p)}_c(t), \hat{\eta}^{(p)}_c(t') \right\} \bigr\rangle &= 
              \frac{\gamma \hbar\omega}{g}\coth\left(\frac{\hbar \omega}{2T}\right)\, \delta(t-t') \label{eq:dynCavd}\,, \\
\bigl\langle \left\{ \hat{\eta}^{(x)}_c(t), \hat{\eta}^{(x)}_c(t') \right\} \bigr\rangle &= 
              \frac{\gamma \hbar g}{\omega}\coth\left(\frac{\hbar \omega}{2T}\right)\, \delta(t-t')\,, \label{eq:dynCave} 
\end{align}
\end{subequations}
with no easily identified classical limit as $\hbar\to 0$. 
Moreover, the reader may have already noticed that the noise-averaged equations of motion~(\ref{eq:dynCava}, \ref{eq:dynCavb}) 
of the cavity model do not satisfy Ehrenfest's theorem~\cite{bondar2016,cohen2018}, 
whereas those of the friction model do, see Eqs.~(\ref{eq:dynBMa}, \ref{eq:dynBMb}). 
Hence the cavity model does not reduce to the Brownian motion in the classical limit $\hbar\to 0$~\cite{lindblad1976brownian,kohen1997}.

In spite of the apparent differences, both noise correlators (\ref{eq:dynBMc}, \ref{eq:dynBMd}) and (\ref{eq:dynCavc}, \ref{eq:dynCavd}, \ref{eq:dynCave}) 
can be derived from the same phenomenological requirements on observables~\cite{araujo2019axiomatic}:
\begin{itemize}
    \item[(A)] the canonical equal-time commutator $\bigl\langle\bigl[ \hat{x}(t),\hat{p}(t)\bigr]\bigr\rangle=\II\hbar$. This avoids that the average commutator could decay to zero; 
    \item[(B)] the Kubo formula of linear response theory;
    \item[(C)] the virial theorem, both classical~\cite{becker2013theorie,collins1978virial,harwit1988astrophysical} 
    and quantum-mechanical~\cite{fock1930bemerkung,araujo2019axiomatic}, in the stationary (${\rm stat}$) long-time limit. For the harmonic oscillator it reads 
    $\bigl\langle \hat{p}^2\bigr\rangle_{\rm stat}-\langle \hat{p}\bigr\rangle_{\rm stat}^2 
    = \frac{\omega^2}{g^2} \left(\bigl\langle\hat{x}^2\bigr\rangle_{\rm stat}-\bigl\langle \hat{x}\bigr\rangle_{\rm stat}^2\right)$. 
    The validity of the virial theorem is a necessary physical criterion for achieving an equilibrium stationary state;
    \item[(D)] the variance at equilibrium for quantum harmonic oscillators, which reads 
    $\bigl\langle\hat{x}^2\bigr\rangle_{\rm stat}-\bigl\langle \hat{x}\bigr\rangle_{\rm stat}^2=\frac{\hbar g}{2\omega}\coth\left(\frac{\hbar\omega}{2 T}\right)$. 
    This requirement forces the evolution to be memoryless and replaces the {\sc qfdt}, which would lead to the correct
    {\em non}-Markovian behaviour~\cite{ford1988,gardiner2004quantum,hanggi2005fundamental,sieberer2015thermodynamic,ford2017fluctuation,araujo2019axiomatic,Weiss2022}.
\end{itemize}
The conditions (A) and (B), imposed for all times $t\geq 0$,
fix the commutators between the noises $\hat{\eta}^{(x)}_\alpha,\hat{\eta}_\alpha^{(p)}$, with $\alpha=f,c$. 
Additionally, the conditions (C) and (D) determine the noise anti-commutators and are designed to drive the stationary state to a quantum equilibrium state.
These phenomenological arguments (A-D) show that there is no evident physical reason why one should prefer one type of dynamics over the other. 
However, we shall see in Sec.~\ref{sec:mast} that the cavity model leads to a {\sc cp} Lindblad generator, whereas the master equation of the friction model is manifestly not 
{\sc cp} but guaranteed to be positive for large enough times.

The requirements (A–D) can be formulated in a compact form by defining the averages
\begin{align} 
C_{-}^{(\mathcal{A},\mathcal{B})}(t,t') &:= \demi \bigl\langle \left[ \hat{\mathcal{A}}(t), \hat{\mathcal{B}}(t') \right]\bigr\rangle \,,\label{eq:Comm2} \\
C_{+}^{(\mathcal{A},\mathcal{B})}(t,t') &:= \demi \bigl\langle \left\{ \hat{\mathcal{A}}(t), \hat{\mathcal{B}}(t') \right\}\bigr\rangle\,.
\label{eq:Comm}
\end{align}
for any operator $\hat{\mathcal{A}}$ and $\hat{\mathcal{B}}$. Therefore, the canonical equal-time commutator (A) reads
\begin{equation}\label{(A)}
    2 \,C_{-}^{(x,p)}(t,t)=\II \hbar\,,
\end{equation}
while the Kubo formula (B) is, see e.g. Ref.~\cite{10.1007/978-3-540-44835-8_7,livi2017nonequilibrium}, 
\begin{equation}
    R^{(x)}(t,s) \,=\, \frac{2\II}{\hbar} \Theta(t-s)\, C_{-}^{(x,x)}(t,s)\,,
\end{equation}
where causality is expressed through the Heaviside function $\Theta$ and
\begin{equation}\label{eq:reponse}
R^{(x)}(t,s) := \left. \frac{\delta \langle\hat{x}(t)\rangle}{\delta \mathscr{B}(s)}\right|_{\mathscr{B}=0}\,,\qquad
R^{(p)}(t,s) := \left. \frac{\delta \langle\hat{p}(t)\rangle}{\delta \mathscr{B}(s)}\right|_{\mathscr{B}=0}\,,
\end{equation} 
are the response functions of the position and momentum variables. 
Herein, $\mathscr{B}(s)$ denotes an {\it a priori} time-dependent magnetic field conjugate to $\hat{x}$. 
On the other hand, one can introduce the connected correlators 
\begin{align}
C_{+,{\rm c}}^{(x,x)}(t,t')&:= C_{+}^{(x,x)}(t,t') - \left\langle \hat{x}(t)\right\rangle\left\langle \hat{x}(t')\right\rangle\,,\label{connected_x}\\
C_{+,{\rm c}}^{(p,p)}(t,t')&:= C_{+}^{(p,p)}(t,t') - \left\langle \hat{p}(t)\right\rangle\left\langle \hat{p}(t')\right\rangle\,,\label{connected_p}
\end{align}
such that the virial theorem reads (C) 
\begin{equation}
    C_{+,{\rm c},{\rm stat}}^{(p,p)} = \frac{\omega^2}{g^2} C_{+,{\rm c},{\rm stat}}^{(x,x)}
\end{equation}
and the variance at equilibrium (D) becomes, as expected,
\begin{equation}
    C_{+,{\rm c, stat}}^{(x,x)} = \frac{\hbar g}{2\omega}\coth\left(\frac{\hbar\omega}{2 T}\right)\,.
\end{equation}

\subsection{Friction model}
Inserting the Hamiltonian~(\ref{eq:H}) into the Eqs.~(\ref{eq:dynBMa}, \ref{eq:dynBMb}) of the friction model, we get 
\begin{align}
\partial_t \hat{x} \,=&\, g \hat{p} +\hat{\eta}_f^{(x)}\,,\\
\partial_t \hat{p} \,=&\, -\frac{\omega^2}{g} \hat{x} - \gamma \hat{p} + B + \hat{\eta}_f^{(p)}\,,
\end{align}
which agree with those of the Redfield model~\cite{kohen1997,Weiss2022}. 
Their formal solution is straightforward and reads,
\begin{align} 
 \hat{x}(t) =&\, \hat{x}_+(0) \e^{-\Lambda_+ t} + \hat{x}_{-}(0) \e^{-\Lambda_{-} t} -\frac{1}{\Lambda_+ - \Lambda_-} 
      \int_0^t \!\D\tau\: \e^{-\Lambda_{+} (t-\tau)} \left[ g\bigl( \hat{\eta}_f^{(p)}(\tau) +B \bigr) + \Lambda_{-}\hat{\eta}_f^{(x)}(\tau) \right]\nonumber \\
      & +\frac{1}{\Lambda_+ - \Lambda_-} 
      \int_0^t \!\D\tau\: \e^{-\Lambda_{-} (t-\tau)} \left[ g\bigl( \hat{\eta}_f^{(p)}(\tau)+ B\bigr) + \Lambda_{+}\hat{\eta}_f^{(x)}(\tau) \right]\,,
      \label{eq:2.9a} \\
      \nonumber\\
 \hat{p}(t) =&\, -\frac{\Lambda_+}{g} \hat{x}_+(0) \e^{-\Lambda_+ t} -\frac{\Lambda_{-}}{g}  \hat{x}_{-}(0) \e^{-\Lambda_{-} t} +\frac{\Lambda_{+}}{g(\Lambda_+ - \Lambda_-)} 
      \int_0^t \!\D\tau\: \e^{-\Lambda_{+} (t-\tau)} \left[ g\bigl( \hat{\eta}_f^{(p)}(\tau) +B\bigr) +  \Lambda_{-}\hat{\eta}_f^{(x)}(\tau) \right]\nonumber \\
      &-\frac{\Lambda_{-}}{g(\Lambda_+ - \Lambda_-)} 
      \int_0^t \!\D\tau\: \e^{-\Lambda_{-} (t-\tau)} \left[ g\bigl( \hat{\eta}_f^{(p)}(\tau) + B\bigr) + \Lambda_{+}\hat{\eta}_f^{(x)}(\tau) \right]\,,
      \label{eq:2.9b} 
\end{align}
using the eigenvalues $\Lambda_{\pm} = \frac{\gamma}{2} \pm \sqrt {\frac{\gamma^2}{4} -\omega^2\,}$ 
and the initial operator values $\hat{x}_{\pm}(0)$. Later, we shall distinguish the cases where $\Lambda_{\pm}$ are real or complex, leading to over-damped and under-damped behavior.
As explained in Ref.~\cite{araujo2019axiomatic}, using the solutions~(\ref{eq:2.9a},~\ref{eq:2.9b}) 
it can be shown that the conditions (A-D) imply the second moments~(\ref{eq:dynBMc},~\ref{eq:dynBMd}) 
of the noises, which are also identical to the ones proposed by Bedeaux and Mazur~\cite{bedeaux2001mesoscopic,bedeaux2002non}. 

As the Lindblad equations are conveniently formulated in terms of the
lowering and raising operators~(\ref{eq:aa}), we now provide the Langevin equation for these operators
\begin{equation}\label{eq:2.12a}
    \partial_t \hat{a} = - \left( \II \omega + \frac{\gamma}{2} \right) \hat{a} 
                       + \frac{\gamma}{2} \hat{a}^{\dag} + \II B \sqrt{\frac{g}{2\hbar\omega}\,} + \hat{\eta}^{(a)}_f\,,
\end{equation}
with the centred noise operator
\begin{equation}\label{eq:2.14a}
    \hat{\eta}^{(a)}_f := \sqrt{\frac{\omega}{2\hbar g}\,} \hat{\eta}^{(x)}_f + \II \sqrt{\frac{g}{2\hbar\omega}\,} \hat{\eta}^{(p)}_f\,,
\end{equation}
and the equation for $\hat{a}^{\dag}$ obtained by formal complex conjugation of Eq.~\eqref{eq:2.12a}. The noise operator $\hat{\eta}^{(a)}_f$ is subject to the non-vanishing second moments 
\begin{subequations}
\begin{align}
\bigl\langle \left[ \hat{\eta}^{(a)}_f(t), \hat{\eta}^{(a^{\dag})}_f(t') \right] \rangle &= \gamma\, \delta(t-t')\,, \\
\bigl\langle \left\{ \hat{\eta}^{(a)}_f(t), \hat{\eta}^{(a^{\dag})}_f(t') \right\} \rangle &= 
              \gamma \coth\left(\frac{\hbar \omega}{2T}\right)\, \delta(t-t')\,, \\
\bigl\langle \left\{ \hat{\eta}^{(a)}_f(t), \hat{\eta}^{(a)}_f(t') \right\} \bigr\rangle 
&= \bigl\langle \left\{ \hat{\eta}^{(a^{\dag})}_f(t), \hat{\eta}^{(a^{\dag})}_f(t') \right\} \bigr\rangle = -\gamma \coth\left(\frac{\hbar\omega}{2T}\right)\, \delta(t-t')\,,
\end{align}
\end{subequations}
where the Hermitian conjugate $\hat{\eta}^{(a^{\dag})}_f=\mbox{$\hat{\eta}_f^{(a)}$}^{\dag}$ is obtained from Eq.~(\ref{eq:2.14a}). This leads to the equations of motion 
\begin{subequations} \label{eq:fric-a}
\begin{align}
\partial_t \bigl\langle \hat{a} \bigr\rangle =& - \left( \II \omega + \frac{\gamma}{2} \right) \bigl\langle\hat{a} \bigr\rangle
                       + \frac{\gamma}{2} \bigr\langle\hat{a}^{\dag}\bigr\rangle + \II B \sqrt{\frac{g}{2\hbar\omega}\,}\,,
    \label{eqF1}\\
    \partial_t \bigl\langle \hat{a} \hat{a} \bigr\rangle =& - 2 \left( \II \omega + \frac{\gamma}{2}\right) \bigl\langle \hat{a}\hat{a} \bigr\rangle 
       + \gamma \bigl\langle \hat{a}^{\dag} \hat{a}\bigr\rangle +\II B \sqrt{\frac{2g}{\hbar\omega}\,}
       \bigl\langle \hat{a} \bigr\rangle- \frac{\gamma}{2} \left( \coth\left( \frac{\hbar\omega}{2T} \right) -1 \right)\,,
    \label{eqF2}\\
        \partial_t \bigl\langle \hat{a}^{\dag} \hat{a} \bigr\rangle =& - \gamma \bigl\langle \hat{a}^{\dag} \hat{a} \bigr\rangle
       + \frac{\gamma}{2} \left( \bigl\langle \hat{a} \hat{a} \bigr\rangle 
       + \bigl\langle \hat{a}^{\dag} \hat{a}^{\dag} \bigr\rangle \right) 
       - \II B \sqrt{\frac{g}{2\hbar\omega}\,}\left(\bigl\langle \hat{a} \bigr\rangle - \bigl\langle \hat{a}^{\dag} \bigr\rangle\right)
       + \frac{\gamma}{2} \left( \coth\left( \frac{\hbar\omega}{2T} \right) -1 \right)\,,
    \label{eqF3}
\end{align}
\end{subequations}
that can be solved by standard techniques. We shall use them in Sec.~\ref{sec:mast} to derive the equivalent master equation and in 
Sec.~\ref{sec:mean} to discuss a physical interpretation of the equilibrium solution.

\subsection{Cavity model}
Using the Hamiltonian~(\ref{eq:H}) into the Eqs.~(\ref{eq:dynCava}, \ref{eq:dynCavb}) of the cavity model, we get
\begin{align}
\partial_t \hat{x} \, =&\, g \hat{p} - \frac{\gamma}{2} \hat{x} +\hat{\eta}_c^{(x)}\,,\label{eq:2.10a}\\
\partial_t \hat{p} \,=&\, -\frac{\omega^2}{g} \hat{x} - \frac{\gamma}{2} \hat{p} + B + \hat{\eta}_c^{(p)}\,.\label{eq:2.10b}
\end{align}
Such equations are usually obtained from a rotating-wave approximation in the Redfield model, e.g., see Ref.~\cite{kohen1997}, 
and break spatial translation-invariance \cite{kohen1997}. Their formal solution reads 
\begin{align} 
 \hat{x}(t) =&\, \hat{x}_+(0) \e^{-\Lambda_+ t} + \hat{x}_{-}(0) \e^{-\Lambda_{-} t} 
    + \demi \int_0^t \!\D\tau\: \e^{-\Lambda_{+} (t-\tau)} \left[ \hat{\eta}_c^{(x)}(\tau) + \frac{\II g}{\omega} \left( \hat{\eta}_c^{(p)}(\tau)+B\right)  \right]\nonumber \\
      &+\demi \int_0^t \!\D\tau\: \e^{-\Lambda_{-} (t-\tau)}  \left[ \hat{\eta}_c^{(x)}(\tau) - \frac{\II g}{\omega} \left( \hat{\eta}_c^{(p)}(\tau) +B\right) \right]\,,
      \label{eq:2.11a}  \\
      &\nonumber \\
 \hat{p}(t) =&\, -\frac{\II\omega}{g} \hat{x}_+(0) \e^{-\Lambda_+ t} +\frac{\II\omega}{g}  \hat{x}_{-}(0) \e^{-\Lambda_{-} t}- \frac{\II\omega}{2g} \int_0^t \!\D\tau\: 
             \e^{-\Lambda_{+} (t-\tau)} \left[ \hat{\eta}_c^{(x)}(\tau) + \frac{\II g}{\omega}\left( \hat{\eta}_c^{(p)}(\tau)+B\right) \right]\nonumber\\
     & + \frac{\II\omega}{2g} \int_0^t \!\D\tau\: \e^{-\Lambda_{-} (t-\tau)} 
           \left[ \hat{\eta}_c^{(x)}(\tau) - \frac{\II g}{\omega}\left( \hat{\eta}_c^{(p)}(\tau)+B\right) \right]\,,
      \label{eq:2.11b} 
\end{align}
with the eigenvalues $\Lambda_{\pm} = \frac{\gamma}{2} \pm \II \omega$.\footnote{In the limit $\omega\gg \gamma$, 
the friction model eigenvalues $\Lambda_{\pm}$ reduce to the ones of the cavity model, leading in that limit to an analogous behaviour.} 
In~\ref{appendixA}, we show that the conditions (A-D) now imply the second moments (\ref{eq:dynCavc}, \ref{eq:dynCavd}, \ref{eq:dynCave}) of the noises. 
Moreover, one may observe that the relaxation rates $\Lambda_{\pm}$ of the cavity model are complex conjugates of each other, leading exclusively to the under-damped regime.

Again, as the Lindblad equations are conveniently formulated in terms of the
lowering and raising operators~(\ref{eq:aa}), we now provide the Langevin equation for these operators
\begin{equation}\label{eq:2.12b}
\partial_t \hat{a} = - \left( \II \omega + \frac{\gamma}{2} \right) \hat{a} 
                       + \II B \sqrt{\frac{g}{2\hbar\omega}\,} + \hat{\eta}^{(a)}_c\,,
\end{equation}
with the centred noise operator
\begin{equation}\label{eq:2.14b}
    \hat{\eta}^{(a)}_c := \sqrt{\frac{\omega}{2\hbar g}\,} \hat{\eta}^{(x)}_c + \II \sqrt{\frac{g}{2\hbar\omega}\,} \hat{\eta}^{(p)}_c\,,
\end{equation}
and the equation for $\hat{a}^{\dag}$ obtained by formal complex conjugation of Eq.~\eqref{eq:2.12b}. The noise operator $\hat{\eta}^{(a)}_c$ is subject to the non-vanishing second moments 
\begin{subequations}
\begin{align}
\bigl\langle \left[ \hat{\eta}^{(a)}_c(t), \hat{\eta}^{(a^{\dag})}_c(t') \right] \rangle &= \gamma\, \delta(t-t')\,, \\
\bigl\langle \left\{ \hat{\eta}^{(a)}_c(t), \hat{\eta}^{(a^{\dag})}_c(t') \right\} \rangle &= 
              \gamma \coth\left(\frac{\hbar \omega}{2T}\right)\, \delta(t-t')\,,
\end{align}
\end{subequations}
where the Hermitian conjugate $\hat{\eta}^{(a^{\dag})}_c=\mbox{$\hat{\eta}_c^{(a)}$}^{\dag}$ is obtained from Eq.~(\ref{eq:2.14b}). This leads to the equations of motion 
\begin{subequations} \label{eq:cavi-a} 
\begin{align}
\partial_t \bigl\langle \hat{a} \bigr\rangle =& - \left( \II \omega + \frac{\gamma}{2} \right) \bigl\langle \hat{a} \bigr\rangle
                       + \II B \sqrt{\frac{g}{2\hbar\omega}\,}\,,
    \label{eqC1}\\
    \partial_t \bigl\langle \hat{a} \hat{a} \bigr\rangle =& - 2 \left( \II \omega 
    + \frac{\gamma}{2}\right) \bigl\langle \hat{a}\hat{a} \bigr\rangle  +\II B \sqrt{\frac{2g}{\hbar\omega}\,}\bigl\langle \hat{a} \bigr\rangle \,,
    \label{eqC2}\\
    \partial_t \bigl\langle \hat{a}^{\dag} \hat{a} \bigr\rangle =& - \gamma \bigl\langle \hat{a}^{\dag} \hat{a} \bigr\rangle
        - \II B \sqrt{\frac{g}{2\hbar\omega}\,}\left(\bigl\langle \hat{a} \bigr\rangle - \bigl\langle \hat{a}^{\dag} \bigr\rangle\right)
       + \frac{\gamma}{2} \left( \coth\left( \frac{\hbar\omega}{2T} \right) -1 \right)\,,
    \label{eqC3}
\end{align}
\end{subequations}
that can be solved by standard techniques. They will be used in Sec.~\ref{sec:mast} to derive the equivalent Lindblad equation and in 
Sec.~\ref{sec:mean} to discuss their stationary solution.

\section{Equivalent master equations}
\label{sec:mast}
In this section, we formulate the effective master equations $\partial_t\hat{\rho}={\cal L}(\hat{\rho})$ for the friction and cavity models. 
This furnishes an equivalent formalism, based on a Lindblad-type deterministic master equation, through tracing out the degrees of freedom of the thermal bath. 
In particular, this establishes the equivalence of the cavity and friction models, respectively, as defined in Sec.~\ref{sec:mod} via their Langevin equations, with the 
two main classes of quantum master equations analysed in detail in Ref.~\cite{kohen1997}. 
As a result, we show that both models are described each by a dynamical generator which is quadratic in the bosonic creation and annihilation operators. 
Such a dynamics may be solved by using the Wigner picture, as will be done in 
Sec.~\ref{sec:dyn}. 

\subsection{Friction model}

We now seek a Lindblad-type master equation for the friction model. Let $\hat\rho_{f}$ be the reduced density operator of the friction model. 
In view of Lindblad's theorem~\cite{lindblad1976generators}, the dynamics of $\hat\rho_{f}$ is cast into the form
\begin{equation} \label{eq:3.2}
    \partial_t \hat{\rho}_{f} =\mathcal{L}(\hat{\rho}_f)=-\frac{\II}{\hbar}\left[\hat{H}_f,\hat{\rho}_f\right] +\mathcal{D}(\hat{\rho}_f)\,,
\end{equation}
where $\hat{H}_f$ is an effective Hamiltonian and
\begin{equation}\label{general_form}
\mathcal{D}(\hat{\rho}_f)=\sum_{j=1,2}\vep_j D_{\rho_f}(\hat V_j,\hat V_j^\dagger)\,,
\end{equation}
is the dissipator, for some real coefficients $\vep_j$ and Hilbert operators $\hat V_j$ to be determined. Here and below we use the super-operator 
\begin{equation} \label{eq:superop}
D_{\mathcal{C}}\bigl(\hat{\mathcal{A}}\,,\hat{\mathcal{B}}\bigr)=\hat{\mathcal{A}}\,\hat{\mathcal{C}}\,\hat{\mathcal{B}}
    -\frac{1}{2}\left\{\hat{\mathcal{B}}\,\hat{\mathcal{A}},\,\hat{\mathcal{C}}\right\}\,,
\end{equation}
for any Hilbert operators $\hat{\mathcal{A}},\,\hat{\mathcal{B}},\,\hat{\mathcal{C}}$. Without loss of generality, we can restrict our study to $\vep_j =\pm 1$, 
and we recover the Lindblad evolution if $\vep_1 =\vep_2 = 1$ \cite{lindblad1976generators}. 
In order to make the master equation~\eqref{eq:3.2} consistent with the Langevin equations~(\ref{eq:dynBMa}, \ref{eq:dynBMb}), we proceed in
two steps: first, we construct the effective Hamiltonian $\hat{H}_f$; second we derive the dissipator $\mathcal{D}(\hat{\rho}_f)$, 
showing that $\vep_j$ must be chosen appropriately to match the Langevin dynamics~(\ref{eq:dynBM}).

Following Refs.~\cite{lindblad1976brownian,isar1994open}, the dissipation operators $\hat{V}_j$ must be linear functions of $\hat{x}$ and $\hat{p}$, 
as the space of the first-degree polynomials spanned by position and momentum must be closed under the action of the Liouvillian~\eqref{eq:3.2}. This suggests the 
ansatz~\cite{lindblad1976brownian,isar2006damped,isar1994open,isar2003lindblad}
\begin{equation}
\hat{V}_j = \mathfrak{a}_j\, \hat{x} + \mathfrak{b}_j\, \hat{p} 
= \alpha_j\, \hat{a} + \beta_j^*\, \hat{a}^{\dag}\,, \qquad \forall j=1,2   
\end{equation}
where the complex numbers $\alpha_j,\beta_j$ (or equivalently $\mathfrak{a}_j, \mathfrak{b}_j$)
are to be found. In general, the operators $\hat{V}_j$ are defined up to constant additive terms, which can always be re-absorbed into the effective Hamiltonian $\hat{H}_f$. 
Therefore, the equations of motion of the one-point functions $\bigr\langle\hat{a}\bigr\rangle$ and $\bigr\langle\hat{a}^\dag\bigr\rangle$ 
come out from Eq.~(\ref{eq:3.2}) if the effective Hamiltonian is taken as~\cite{lindblad1976brownian,lindblad1976generators,isar1994open,isar2003lindblad,isar2006damped}
\begin{equation} \label{eq:3.5}
\hat H_f = \hat H + \hat H_{\gamma} = \hat H + \frac{\gamma}{4} \left( \hat{p}\hat{x} + \hat{x}\hat{p} \right) 
= \hat H + \frac{\II\hbar}{4} \gamma \left( \hat{a}^{\dag}\hat{a}^{\dag} - \hat{a}\hat{a}\right)\,.
\end{equation}
In contrast to the often-considered {\em Lamb shift}~\cite{breuer2002theory}, 
the additional contribution $\hat H_{\gamma}$ does not commute with the system Hamiltonian $\hat H$. As reviewed in detail in Ref.~\cite{isar1994open}, we define
\begin{equation}\label{eq:Edef}
E_1 := \frac{1}{\hbar} \sum_{j=1,2} \vep_j |\alpha_j|^2 \,,\qquad
E_2 := \frac{1}{\hbar} \sum_{j=1,2} \vep_j |\beta_j|^2 \,,\qquad
E_3 := \frac{1}{\hbar} \sum_{j=1,2} \vep_j \alpha_j \beta_j = E_4^*   \,, 
\end{equation}
which depend implicitly on the Hamiltonian's parameters, and $\gamma$. The master equation~(\ref{eq:3.2}) with the effective Hamiltonian~(\ref{eq:3.5}) takes the form
\begin{align}
 \frac{\D\hat{\rho}_f}{\D t}\;=&\;\; 
E_1 \hat{a}\hat{\rho}_f\hat{a}^{\dag} - \demi\bigl( E_1 +\II\omega\bigr) \hat{a}^{\dag}\hat{a}\hat{\rho}_f 
                                      - \demi\bigl( E_1 -\II\omega\bigr) \hat{\rho}_f\hat{a}^{\dag}\hat{a}
+ E_2 \hat{a}^{\dag}\hat{\rho}_f\hat{a} \nonumber \\
 &\;\;
- \demi\bigl( E_2 +\II\omega\bigr) \hat{a}\hat{a}^{\dag}\hat{\rho}_f 
                                        - \demi\bigl( E_2 -\II\omega\bigr) \hat{\rho}_f\hat{a}\hat{a}^{\dag} 
 + E_3 \hat{a}\hat{\rho}_f\hat{a} - \demi\left( E_3 +\frac{\gamma}{2}\right) \hat{a}\hat{a}\hat{\rho}_f \nonumber\\
 &\;\;
                                     - \demi\left( E_3 -\frac{\gamma}{2}\right) \hat{\rho}_f\hat{a}\hat{a}
+ E_4 \hat{a}^{\dag}\hat{\rho}_f\hat{a}^{\dag} - \demi\left( E_4 -\frac{\gamma}{2}\right) \hat{a}^{\dag}\hat{a}^{\dag}\hat{\rho}_f 
                                           - \demi\left( E_4 +\frac{\gamma}{2}\right) \hat{\rho}_f\hat{a}^{\dag}\hat{a}^{\dag} \,.
\label{eq:3.8}
\end{align}
The second step, namely the determination of $E_{\ell}$ ($\ell=1,\ldots,4$), is achieved by comparing Eq.~\eqref{eq:3.8} 
with the requested equations of motion~(\ref{eq:fric-a}). 
Using the ansatz~\eqref{eq:3.8}, the single-point function obeys
\begin{equation}\label{eq:3.9}
\frac{\D \langle\hat{a}\rangle}{\D t}
= - \left( \demi \bigl( E_1 - E_2\bigr) +\II\omega \right) \bigl\langle \hat{a}\bigr\rangle 
  +\frac{\gamma}{2} \bigl\langle \hat{a}^{\dag} \bigr\rangle\,,
\end{equation}
which reproduces Eq.~(\ref{eqF1}) if 
\begin{align} \label{eq:Ela}
E_1 - E_2\stackrel{!}{=}\gamma\,.
\end{align}
Since the equation of motion for $\langle\hat{a}^{\dag}\rangle$ is the Hermitian conjugate of Eq.~(\ref{eq:3.9}), it does not yield any further information. 
Next, the average $\langle \hat{a}\hat{a}\rangle$ obeys
\begin{equation}\label{eq:3.10}
    \frac{\D \langle\hat{a}\hat{a}\rangle}{\D t} = - \bigl( E_1 - E_2+2\II\omega\bigr)  \bigl\langle \hat{a}\hat{a}\bigr\rangle 
+ {\gamma} \bigl\langle \hat{a}^{\dag} \hat{a}\bigr\rangle - \left( E_4 -\frac{\gamma}{2} \right)\,,
\end{equation}
such that comparison with Eq.~(\ref{eqF2}), along with Eq.~(\ref{eq:Ela}), gives
\begin{equation}\label{eq:Elb}
E_4 \stackrel{!}{=} E_3 = \frac{\gamma}{2} \coth \left( \frac{\hbar \omega}{2 T} \right) = \gamma\bigl( n_{\omega}+1/2\bigr)    \,,
\end{equation} 
with $n_{\omega}=\bigl[ e^{\hbar\omega/T}-1\bigr]^{-1}>0$ being the Bose-Einstein distribution. Again, $\langle \hat{a}^{\dag}\hat{a}^{\dag}\rangle =\langle \hat{a}\hat{a}\rangle^*$ 
does not yield any further information. Finally, the average $\langle \hat{a}^{\dag}\hat{a}\rangle$ obeys
\begin{equation}\label{eq:3.11}
\frac{\D \langle\hat{a}^{\dag}\hat{a}\rangle}{\D t} = - \bigl( E_1 - E_2\bigr)  \bigl\langle \hat{a}^{\dag}\hat{a}\bigr\rangle 
+ \frac{\gamma}{2} \bigl\langle \hat{a}^{\dag} \hat{a}^{\dag} +  \hat{a} \hat{a}\bigr\rangle + E_2     \,,
\end{equation}
such that comparison with Eq.~\eqref{eqF3} and using once more Eq.~\eqref{eq:Ela} leads to 
\begin{equation} \label{eq:Elc}
E_2 \stackrel{!}{=} \frac{\gamma}{2} \left( \coth \left( \frac{\hbar \omega}{2 T} \right) -1 \right) = \gamma n_{\omega}\,.
\end{equation}
Combining Eq.~\eqref{eq:Ela} with Eq.~\eqref{eq:Elc}, we finally get
\begin{equation}\label{eq:Ela2}
    E_1 = \frac{\gamma}{2} \left( \coth \left( \frac{\hbar \omega}{2 T} \right) +1 \right)
= \gamma \bigl( n_{\omega} + 1 \bigr)\,.
\end{equation}
Therefore, the equivalence between the Langevin equations (\ref{eq:dynBM}) and the Lindblad-type equation (\ref{eq:3.2})  
is established through the identical equations of motion, as required. The full Liouvillian takes the form
\begin{align} \label{eq:3.13} 
 \mathcal{L}(\hat{\rho}_f)=&-\frac{\II}{\hbar}\left[\hat{H}_f,\hat{\rho}_f\right] 
 +\gamma \bigl( n_{\omega} + 1 \bigr)D_{\rho_f}(\hat a,\hat a^\dagger)     + \gamma n_{\omega}D_{\rho_f}(\hat a^\dag,\hat a) \nonumber \\
 &+ \gamma \left( n_{\omega} + \frac{1}{2} \right)D_{\rho_f}(\hat a,\hat a) + \gamma \left( n_{\omega} + \frac{1}{2} \right)D_{\rho_f}(\hat a^\dag,\hat a^\dag)\,.
\end{align}
If $T\gg \hbar\omega$, Eq.~\eqref{eq:3.13} reproduces the equation of motion for $\hat{\rho}_f$ derived from the Caldeira-Leggett model \cite{CaldeiraLegget1983,kohen1997,Rama2009}. 
Taking this limit is required when (\ref{eq:3.13}) is derived from an explicit microscopic model of the bath, but this condition does not seem to be necessary in our purely phenomenological approach. 

The question of how to invert Eq.~\eqref{eq:Edef} to determine the constants $\alpha_j$ and $\beta_j$ (for $j=1,2$) 
has not yet been addressed in the literature. Remarkably, it turns
out that this inversion is impossible if 
$\vep_1=\vep_2=1$.\footnote{Eqs.~(\ref{eq:Elb},~\ref{eq:Elc},~\ref{eq:Ela2}) imply $E_3 = E_4^* = \demi \bigl( E_1 + E_2\bigr)$. 
Combining the latter identity with the assumption $\vep_1=\vep_2=1$,
\[
\sum_{j=1,2} \bigl| \alpha_j - \beta_j^* \bigr|^2 = 0\,,
\]
which implies $\alpha_1 = \beta_1^*$ and $\alpha_2 = \beta_2^*$, and thus $E_1=E_2$ by Eq.~\eqref{eq:Edef}. 
However, the equality $E_1=E_2$ contradicts Eq.~\eqref{eq:Ela} for $\gamma>0$. 
We conclude that no solution exists under the assumption $\vep_1=\vep_2=1$, which is therefore inadmissible. 
Consequently, the friction model is not determined by a Lindblad evolution.}
For a physical solution of the coefficients~(\ref{eq:Edef}), we must rather take $\vep_1=+1$ and $\vep_2=-1$, and together with Eqs.~(\ref{eq:Elb}, \ref{eq:Elc}, \ref{eq:Ela2}) we get
\begin{equation}
    \alpha_1^2-\alpha_2^2=\gamma (n_\omega+1)\,,
\end{equation}
\begin{equation}
    \beta_1^2-\beta_2^2=\gamma n_\omega \,, 
\end{equation}
\begin{equation}
    \alpha_1\beta_1-\alpha_2\beta_2=\gamma (n_\omega+1/2)\,.
\end{equation}
Since we have three independent equations for four parameters, one of them will remain free and can be chosen for convenience. A simple parametrisation is  
\begin{equation}
\alpha_1=\alpha\cosh\phi\,,\quad \alpha_2=\alpha\sinh\phi\,,\quad\beta_1=\beta\cosh\psi\,,\quad \beta_2=\beta\sinh\psi\,,
\end{equation}
with $\alpha>0$ and $\beta>0$ from which it follows 
\begin{equation}
    \alpha^2=\gamma (n_\omega+1)\,,\quad \beta^2=\gamma n_\omega\,,\quad \cosh(\phi-\psi)=\cosh{\frac{\omega}{2T}}\,.
\end{equation}
We always have $0<\alpha_2 <\alpha_1$ and $0<\beta_2 < \beta_1$, while the parameter $\phi+\psi$ remains arbitrary. Finally, the Liouvillian (\ref{eq:3.13}) becomes
\begin{align}\label{Lind_nonCP}
\mathcal{L}(\hat{\rho}_f) =& -\frac{\II}{\hbar}\bigl[ \hat{H}_f, \hat{\rho}_f\bigr] 
  + D_{\rho_f}\bigl( \alpha \cosh\phi\,\hat{a} +\beta\cosh\psi\, \hat{a}^{\dag}, \alpha \cosh\phi\,\hat{a}^{\dag} +\beta\cosh\psi\, \hat{a}\bigr)\nonumber\\
& - D_{\rho_f}\bigl( \alpha \sinh\phi\,\hat{a} +\beta\sinh\psi\, \hat{a}^{\dag}, \alpha \sinh\phi\,\hat{a}^{\dag} +\beta\sinh\psi\, \hat{a}\bigr) \,.
\end{align}

\subsection{Cavity model}
The cavity model is nothing else than the paradigmatic example of
the dampening of a single electromagnetic field mode inside a cavity~\cite{kohen1997,carmichael1999statistical,breuer2002theory,gardiner2004quantum,schaller2014open,alicki2023}. 
The environment may be provided by the modes outside the cavity and acts as a thermal bath. 
Let $\hat\rho_{c}$ be the reduced density operator of the cavity model. The dynamics of $\hat\rho_{c}$ is provided by the Liouvillian  
$\mathcal{L}(\hat{\rho}_{c})=\partial_t\hat\rho_{c}$, which takes the form
\begin{equation}\label{eq:cavity}
      \partial_t \hat{\rho}_{c}=\mathcal{L}(\hat{\rho}_c)
      =-\frac{\II}{\hbar}\left[\hat{H},\hat{\rho}_c\right] + \gamma(n_\omega+1)D_{\rho_c}(\hat a,\hat a^\dagger) + \gamma n_\omega D_{\rho_c}(\hat a^\dagger,\hat a)\,,
\end{equation}
where 
\begin{equation}\label{eq:cavity_diss}
    \mathcal{D}(\hat{\rho}_{c})=\gamma(n_\omega+1)D_{\rho_c}(\hat a,\hat a^\dagger) + \gamma n_\omega D_{\rho_c}(\hat a^\dagger,\hat a)\,.
\end{equation}
is the Lindblad dissipator which describes single-particle gain and loss processes with jump frequencies 
$\gamma n_\omega$ and $\gamma(n_\omega+1)$. 
Using straightforward algebra, one can verify that the Liouvillian (\ref{eq:cavity})
reproduces the equations of motion~(\ref{eq:cavi-a}) 
thereby establishing the equivalence between the Langevin and Lindblad approaches for the cavity model.

The dynamical map generated by the Liouvillian~\eqref{eq:cavity} is manifestly {\sc cp}, whereas the one corresponding to the friction model~\eqref{Lind_nonCP} is not. 
We remember that, by definition, a quantum dynamical map $\epsilon_t$ is {\sc cp} if the open system (S) can be extended by $n$ non-interacting ancill{\ae}—referred to 
as 'witnesses' (W)—such that the extended map 
$\epsilon_t\otimes\mathbb{I}_n$, where $\epsilon_t$ and $\mathbb{I}_n$ act on the system and ancill{\ae} respectively, preserves the positivity of the joint system-witness density matrix 
$\hat{\rho}_{\rm SW}$ for any $n\in\mathbb{N}$~\cite{lindblad1976generators,gorini1976,breuer2016colloquium}. 
More specifically, it can be shown that complete positivity holds if and
only if the dissipators in the Liouvillian are of the form~\eqref{general_form} where all 
$\vep_j\geq 0$~\cite{lindblad1976generators,gorini1976,breuer2002theory}, which is fulfilled for the dissipator~\eqref{eq:cavity_diss}, but is in contrast with Eq.~\eqref{Lind_nonCP}. Indeed, it is well known that the Liouvillian~\eqref{Lind_nonCP} is not {\sc cp} for the Caldeira-Leggett model~\cite{CaldeiraLegget1983,kohen1997}. 
Alternatively, this could have been inferred from the lack of positivity for sufficiently squeezed initial states for short times \cite{talkner1986}.

Rather than revisiting the longstanding and ongoing debate over the need of complete positivity as a dynamical postulate~\cite{cuffaro2013}, 
we shall focus exclusively on the observable behaviour of the system and refrain from making any specific assumptions about the bath or potential witnesses.
Therefore, we shall compare the formal and physically observable properties of the friction and cavity models, 
whose Langevin noises are derived from the {\em same} physical criteria (A-D) on observable behaviour, 
demonstrating that the friction model—whether or not it satisfies complete positivity—can be physically admissible within specific regimes.

\section{Wigner function dynamics}
\label{sec:dyn}

The phase-space formulation of quantum dynamics offers three key advantages in our analysis. 
First, it enables a direct comparison with the detailed study of quantum master equations in Ref.~\cite{kohen1997}, formulated in terms of Wigner functions. 
Second, the Liouvillians in Eqs.~\eqref{Lind_nonCP} and~\eqref{eq:cavity} are mapped to Fokker-Planck-type differential equations for the Wigner functions, 
which can be solved using standard analytical techniques. 
Third, this formulation facilitates the study of positivity, which is a fundamental requirement of quantum mechanics. 
While positivity is guaranteed in the cavity model due to its {\sc cp}-divisible dynamical map, see Eq.~\eqref{eq:cavity_diss}, this is not the case for the friction model, 
where positivity holds only for a restricted set of initial states that are not strongly squeezed~\cite{talkner1986,kohen1997}. 
Since, according to the principles of quantum mechanics, a proper density matrix must exhibit non-negative eigenvalues to conform to the probabilistic framework, 
such classes of initial states are precisely those for which the Markovian approximation should be reliable and physically admissible. 
Nonetheless, as we shall demonstrate, positivity is always restored at long times.
Any failure of positivity at intermediate times, for certain sufficiently strongly squeezed initial conditions \cite{talkner1986}, should be seen as a consequence of the Markov {\em approximation}.

The {\em Wigner function approach} is 
based on a map between the density matrix and a real-valued phase-space function, e.g. 
\cite{weinbub2018recent,hillery1984,moyal1949quantum,lee1995,schleich2001quantum,case2008,hinarejos2012wigner,dean2018wigner,de2021wigner,malouf2019wigner,santos2017wigner,gardiner2004quantum,Weiss2022}, 
\begin{equation}
    \hat{\rho}_\alpha\mapsto 
    \mathcal{W}_\alpha(x,p):=\frac{1}{2\pi\hbar}\int_{\mathbb{R}} \e^{{\rm i}\,pw/\hbar}\hspace{0.1cm}
                            \langle x-w/2\hspace{0.1cm}|\hspace{0.1cm}\hat{\rho}_\alpha\hspace{0.1cm}|\hspace{0.1cm}x+w/2\rangle\hspace{0.1cm} \D w .
\end{equation}
with $\alpha=f,c$ for friction ($f$) and cavity ($c$) model. $\mathcal{W}_\alpha(t,x,p)$ is the so-called {\em Wigner function} 
and provides a complete description of the system (see~\ref{appendixB} and~\ref{appendixC}  for details and background). 
The expectation value of a quantum observable $\hat{O}$ is given by  
\begin{equation}
    \bigl\langle \hat{O}\bigr\rangle_\alpha = \int_{\mathbb{R}^2} \!\D x\D p\hspace{.1cm} O(x,p)\: \mathcal{W}_\alpha(t,x,p)\,,
\end{equation}
where $O(x,p)$ indicates the Weyl transform of $\hat{O}$, see Tab.~\ref{tab1} for examples.
In~\ref{appendixC}, it is shown that Eqs.~(\ref{Lind_nonCP},~\ref{eq:cavity}) may be recast as linear partial differential equations for the Wigner function,
\begin{equation} \label{W-evo}
   \partial_t \mathcal{W}_\alpha(t,x,p) =\Vec{\nabla} \bigcdot \left[\frac{\Vec{D}_\alpha}{2}\Vec{\nabla}\mathcal{W}_\alpha(t,x,p)+\Vec{\Theta}_\alpha\begin{pmatrix}x \\[-8pt] p
    \end{pmatrix}\mathcal{W}_\alpha(t,x,p)-\Vec{\Theta}_\alpha\Vec{\mu}_\alpha \mathcal{W}_\alpha(t,x,p)\right]\,,
\end{equation}
with $\Vec{\nabla}=\begin{pmatrix} \partial_x \\[-5pt] \partial_p \end{pmatrix}$. The reader is referred to~\ref{appendixC} 
for the detailed notation of the matrices $\Vec{D}_\alpha$ and $\Vec{\Theta}_\alpha$, and the vector $\Vec{\mu}_\alpha$, for $\alpha=f,c$, 
as well as the full analytical solution to the Fokker-Planck equation~\eqref{W-evo}. The reader is also referred to~\ref{appendixD} for the dynamics of Gaussian states. 

In~\ref{appendixC}, it is shown that the relaxation time is expressed through the eigenvalues $\Lambda_{\pm}^{(\alpha)}$ of $\Theta_\alpha$, 
see Eqs.~(\ref{F:eq:Wtilde},~\ref{F:eq:sigma}). For the cavity model, $\Lambda_{\pm}^{(c)}=\gamma/2\pm\II \omega$ 
are always complex, indicating that only an under-damped regime is present. Therefore the characteristic relaxation time is $2/\gamma$, 
since the imaginary part of $\Lambda_{\pm}^{(c)}$ merely generates an oscillatory behaviour. In contrast, in the friction model we distinguish two main scenarios. 
If $\gamma<2\omega$, both eigenvalues $\Lambda_{\pm}^{(f)}=(\gamma\pm\II \sqrt{4\omega^2-\gamma^2})/2$ are complex, corresponding to the under-damped regime, 
with a relaxation time of $2/\gamma$. If $\gamma>2\omega$, both eigenvalues $\Lambda_{\pm}^{(f)}=(\gamma\pm \sqrt{\gamma^2-4\omega^2})/2$ are real, 
distinguishing an under-damped and over-damped regime. This leads to a crossover from rapid short-time decay to slower long-time relaxation, 
characterised by a relaxation time of $2/(\gamma - \sqrt{\gamma^2 - 4\omega^2})$. The case $\gamma = 2\omega$ is analytically straightforward, 
as $\Lambda_{\pm}^{(f)}=\gamma/2$ and again only the under-damped regime shows up. 

As shown in~\ref{appendixE}, in the long-time limit, the system in both models always reaches a formal `equilibrium state' with the bath, i.e. the Gibbs state $\hat{\rho}^{T}_{\alpha}$, 
\begin{equation}\label{thermal_long}
\hat{\rho}^{T}_{\alpha}:=\lim_{t\to\infty} \hat\rho_\alpha(t)=\frac{\e^{-\hat{\mathcal{H}}_{\alpha}/T}}{\tr(\e^{- \hat{\mathcal{H}}_{\alpha}/T})}\,,  
\end{equation}
where
\begin{equation}\label{eq:4.17ham1}
\hat{\mathcal{H}}_f=\hat{H}\,,\qquad
\hat{\mathcal{H}}_c=\hat{H}+B\frac{\gamma^2}{\gamma^2+4\omega^2}\hat{x}-B\frac{2g\gamma}{\gamma^2+4\omega^2}\hat{p}   \,.
\end{equation}
For $B=0$, the Hamiltonians into~(\ref{eq:4.17ham1}) coincide, $\hat{\mathcal{H}}_f=\hat{\mathcal{H}}_c=\hat{H}$, and hence the stationary states are the same in both models, and indeed equilibrium states. 
However, for $B\ne 0$, Eq.~(\ref{eq:4.17ham1}) shows that the effective Hamiltonian $\hat{\mathcal{H}}_c$ has re-entrant  
$\gamma$-dependent contributions. Hence its stationary state does depend on $\gamma$ 
and is not a physical equilibrium state in the usual sense, meaning that the cavity model no longer relaxes to thermal equilibrium. 
This observation will become important in Sec.~\ref{sec:mean}. 
On the other hand, the friction model with effective Hamiltonian $\hat{\mathcal{H}}_f=\hat H$ always guarantees thermalisation, with the associated $\gamma$-independent equilibrium state. 
We point out that the failure of full equilibration arises in the cavity model although the
connected correlators $C_{+,{\rm c},{\rm stat}}^{(x,x)}$ and $C_{+,{\rm c},{\rm stat}}^{(p,p)}$ were required in Sec.~\ref{sec:mod} to take their equilibrium values. 
We emphasize that Eq.~\eqref{thermal_long} also demonstrates that positivity is always restored at long times. However, for the friction model, 
it is less clear whether positivity is preserved throughout the entire time evolution. 
For a detailed analysis of positivity, the reader is referred to \ref{appendixD}, where we show that positivity is preserved during short-time dynamics, 
provided the initial states are not too strongly squeezed~\cite{talkner1986}.

In quantum systems strongly coupled to a reservoir, it is commonly observed that the long-time
limit of quantum dynamics may not relax to stationary states of the Gibbs-Boltzmann form~\cite{Miller2018,Trushechkin2022}. 
While certainly compatible with the general principles of statistical mechanics~\cite{LandauLifschitz1979,PathriaBeale2011}, this effect arises from the fundamentally {\em non}-Markovian quantum dynamics, which introduce significant memory effects through the quantum noise correlators. If the bath is 
modelled by a large set of harmonic oscillators, this derivation is very well-known~\cite{ford1965,ford1988,Weiss2022}. 
The same noise correlators can also be derived from the phenomenological approach of Sec.~\ref{sec:mod} when the requirement (D) is replaced by the {\sc qfdt}~\cite{araujo2019axiomatic}, without making any explicit microscopic hypothesis on the system-bath interaction. 
The stationary state is then described by a {\em Hamiltonian of mean force} ({\sc hmf})~\cite{Miller2018,Trushechkin2022}. 
For a harmonic oscillator with $B=0$, it is well-established that the {\sc hmf} takes the same form as (\ref{eq:H}), with unchanged system variables $\hat{x},\hat{p}$, but with modified
parameters $g=g(\gamma,T)$ and $\omega=\omega(\gamma,T)$~\cite{Grabert1984,HuPazZhang1992,Miller2018}. Here we emphasize that, for the Markovian cavity model with $B\ne 0$, we find the opposite from Eq.~\eqref{eq:4.17ham1}. Indeed, by completing the square (up to additive constants), it is possible to rewrite $\hat{\mathcal{H}}_c$ in the same functional form as in~\eqref{eq:H},
\begin{equation}
   \hat{\mathcal{H}}_c = \frac{g}{2}\hat{P}^2+\frac{\omega^2}{2g}\hat{X}^2 - B \hat{X}\,,
\end{equation}
where $\hat{X} = \hat{x} + \frac{g}{\omega^2} \frac{\gamma^2}{\gamma^2+4\omega^2}B$ and $\hat{P} = \hat{p}-\frac{2\gamma}{\gamma^2+4\omega^2}B$, where $\hat{X}$ and $\hat{P}$ are shifted with respect to the variables $\hat{x},\hat{p}$. These shifts  depend on $B,\gamma,g,\omega$, but not on $T$. On the other hand, the parameters $g,\omega$ in the hamiltonian~\eqref{eq:H} remain unchanged.
Therefore, non-relaxation towards quantum equilibrium of the Markovian cavity model for $B\ne 0$ is not an example of {\sc hmf}.

\section{Effective mean-field theories}
\label{sec:mean}

Mean-field descriptions \cite{nishimori2010elements,henkel2010} 
of phase transitions often provide a first qualitative appreciation of cooperative effects in many-body systems. In their
most simple form, they arise from a replacement of interaction terms by a self-consistently determined external field. In order to further illustrate
the differences between the cavity model and the friction model, 
which are both one-body systems, 
we shall interpret their long-time behaviour as mean-field approximations of a many-body magnet, where the
averaged position $\bigl\langle \hat{x}\bigr\rangle$ becomes the model's magnetisation $m$. 
This is the opposite direction with respect to what is usually carried out in many-body systems, where a self-consistent mean-field  approximation is made in order to replace the
original many-body model by a more simple mean-field theory of a single degree of freedom \cite{nishimori2010elements,henkel2010}. 
As a model of reference, we shall use the the {\em quantum spherical model}, whose equilibrium 
and non-equilibrium critical behaviour is well-understood and shares many qualitative features with more `realistic' models 
\cite{berlin1952spherical,lewis1952spherical,henkel1984,nieuwenhuizen1995quantum,vojta1996quantum,godreche2000response,rokni2004,oliveira2006quantum,sachdev_2011,chandran2013,bienzobaz2013,maraga2015aging,Gagel15,perez2017redfield,barbier2019pre,wald2015spherique,wald2018lindblad,wald2021non}. 
For a recent review and tutorial, see~\cite{henkel2023quantum}.

In order to emphasise quantum effects, we set the bath temperature $T=0$.
Given the single-particle Hamiltonian~\eqref{eq:H2}, 
the spherical constraint $\bigl\langle \hat{x}^2\bigr\rangle=1$ fixes the angular frequency $\omega$. 
In dynamics, $\omega=\omega(t)$ becomes indeed time-dependent, as is well-known in classical dynamics \cite{godreche2000response}. For a single degree of freedom \cite{wald2016lindblad},
\begin{equation} \label{gl:sc}
\omega(t) = \frac{\hbar g}{2} \left(  \bigl\langle \hat{a}^{\dag}\hat{a}^{\dag} \bigr\rangle + \bigl\langle \hat{a}\hat{a}\bigr\rangle 
         + \bigl\langle \hat{a}^{\dag}\hat{a} \bigr\rangle + 1\right)\,.
\end{equation}
The interpretation of this single-body problem as an effective mean-field approximation is completed through a self-consistent form of the magnetic field
\begin{equation} \label{gl:mean}
B = \kappa \bigl\langle x \bigr\rangle = \kappa \sqrt{\frac{\hbar g}{2\omega}\,}\, \bigl\langle \hat{a}^{\dag} + \hat{a}\bigr\rangle\,,
\end{equation}
where $\kappa>0$ is the molecular field constant. For simplicity, we use the following parametrization,
\begin{equation}\label{lista_var}
    \bigl\langle \hat{a}\bigr\rangle = x_1 + \II x_2\,,\qquad 
\bigl\langle \hat{a}\hat{a}\bigr\rangle = x_3 + \II x_4\,,\qquad 
\bigl\langle \hat{a}^{\dag}\hat{a}\bigr\rangle = x_5\,,
\end{equation}
and the equations of motion of the friction and cavity models can be taken from Sec.~\ref{sec:mod}, together with the spherical constraint~(\ref{gl:mean}), which is used to eliminate $x_5$. 

We are interested in the phase diagramme of the stationary state and hereinafter we set $\omega := \lim_{t\to\infty} \omega(t)$. 
The solution of the stationary equations $\D \bigl\langle \hat{a}\bigr\rangle/\D t=\D \bigl\langle \hat{a}\hat{a}\bigr\rangle/\D t=\D \bigl\langle \hat{a}^\dagger\hat{a}\bigr\rangle/\D t=0$ 
is straightforward and we briefly summarize the results below. This study further demonstrates that the cavity model does not describe relaxations to an equilibrium state, 
in contrast to the friction model, as evidenced by the two distinct effective Hamiltonians~(\ref{eq:4.17ham1}) for $B\ne 0$. 

\begin{figure}
    \centering
    \includegraphics[width=0.8\linewidth]{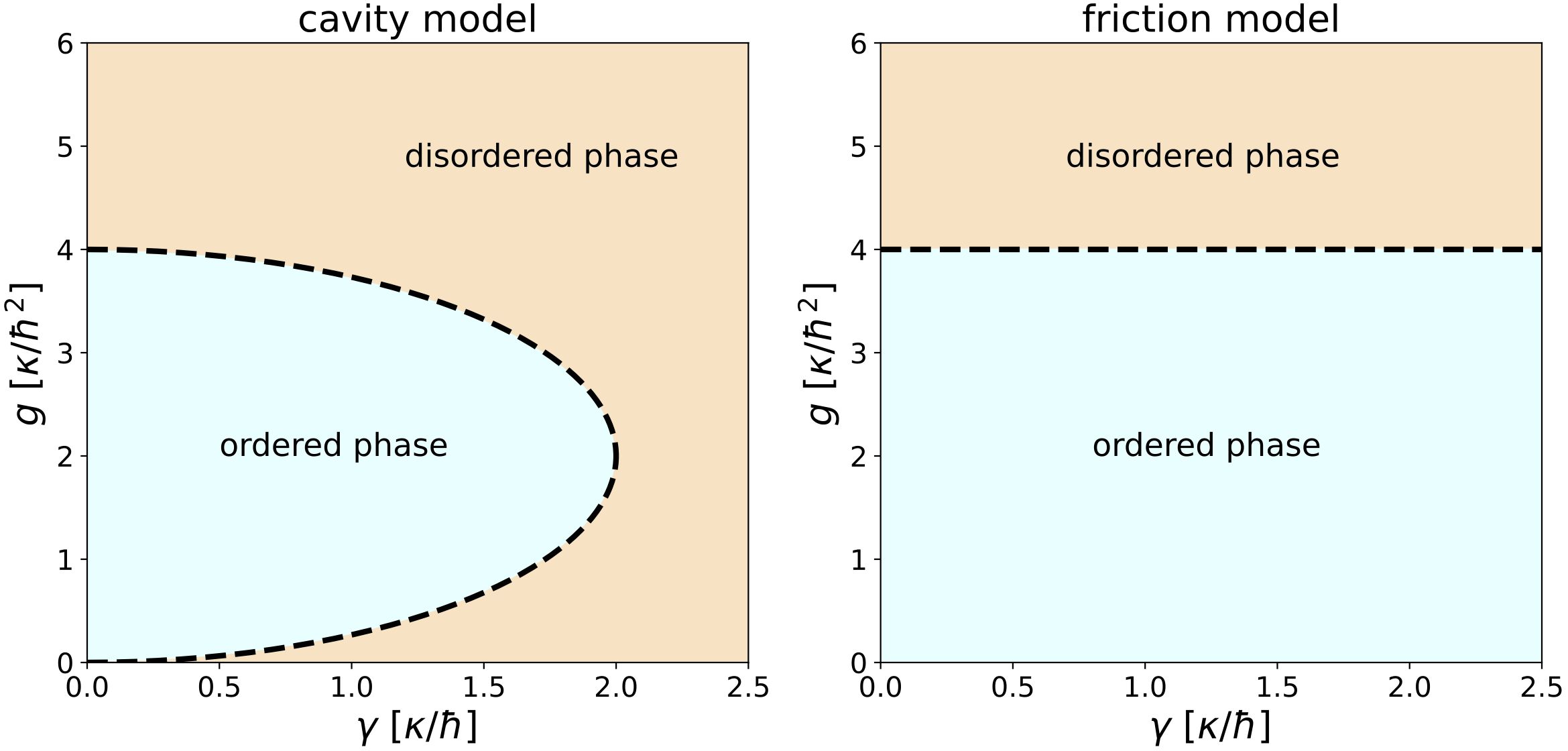}
    \caption[phase-space-diag]{Mean-field phase diagrammes of the quantum spherical model at $T = 0$, for the cavity model (left) and the friction model (right).}
    \label{fig:fig1}
\end{figure}

\subsection{Friction model}
Two distinct stationary solutions are found, both independent of $\gamma$, as required for an equilibrium state. 
The first one corresponds to a disordered, paramagnetic state with $x_1=x_2=x_3=x_4=x_5=0$ and $\omega=\omega_d = {\hbar g}/{2}$, resulting in vanishing magnetisation
$m=0$. The second one corresponds to an ordered, magnetic state with $\omega=\omega_{o}=\sqrt{\kappa g\,}$, $x_2=x_4=0$ and 
from Eqs.~(\ref{eq:fric-a}) together with the definitions (\ref{lista_var}), it  follows straightforwardly that 
\begin{equation}
m^2 = x_1^2 = \frac{1}{4\sqrt{g\,}} \left( \frac{2\sqrt{\kappa\,}}{\hbar} - \sqrt{g\,}\right) = x_3\,.
\end{equation}
At the critical point, $\omega_d=\omega_o$ which gives the critical value $g_c = {4\kappa}/{\hbar}^2$. 
Therefore, the magnetisation becomes $m^2 = \bigl( g_c^{1/2} - g^{1/2}\bigr)/4 g^{1/2}$, with $g_c=4\kappa/\hbar^2$,
such that the system is magnetically ordered for all $g<g_c$, as expected for an ordered magnetic state at equilibrium, see Fig.~\ref{fig:fig1}.  
Since close to criticality, one expects $m^2 \sim \bigl( g_c - g\bigr)^{2\beta}$, one recovers the standard mean-field exponent $\beta=\demi$~\cite{nishimori2010elements,henkel2010}.

\subsection{Cavity model}
Again, from Eqs.~(\ref{eq:cavi-a}) and the definitions (\ref{lista_var}), two distinct stationary solutions are found \cite{wald2016lindblad}. 
The first one is identical to the disordered solution 
$x_1=x_2=x_3=x_4=x_5=0$ and $\omega= \omega_d = {\hbar g}/{2}$
of the friction model. The second solution, however, is now $\gamma$-dependent, 
$\omega=\omega_o' = \sqrt{ g\kappa - \gamma^2/4\,}$ and all variables $x_1,\,x_2,\,x_3,\,x_4$ are non-vanishing. 
The condition of criticality $\omega_d = \omega_o'$ admits two solutions for $\gamma<2\kappa/\hbar$, which leads to two separate critical points~\cite{wald2016lindblad}
\begin{equation}
g_{c,\pm}(\gamma) = \frac{2\kappa}{\hbar^2} \pm \sqrt{ \frac{4\kappa^2}{\hbar^4} - \frac{\gamma^2}{\hbar^2}\,}\,.
\end{equation} 
While in the limit $\gamma\to 0$, 
the upper critical point $\lim_{\gamma\to 0} g_{c,+}(\gamma) = g_c$ goes over to that of the friction model and $\lim_{\gamma\to 0} g_{c,-}(\gamma) = 0$, for a finite 
$\gamma<2\kappa/\hbar$, there is also a lower
critical point $g_{c,-}(\gamma)$, as shown in Fig.~\ref{fig:fig1}. The magnetisation is non-vanishing only for $g_{c,-}(\gamma)<g<g_{c,+}(\gamma)$ and reads~\cite{wald2016lindblad}
\begin{equation}
m^2 = x_1^2 = \frac{\gamma^2}{4\kappa g} \left( 1 + \frac{4\omega^2}{\gamma}\right)
\left( 1 - \frac{\hbar g}{2\omega}\right) \,.
\end{equation}
These re-entrant transitions are a purely kinetic effect and cannot be properties of an equilibrium state. 
Through the value of the damping constant, the cavity strongly influences the behaviour even of the stationary state. 

\section{Discussion and conclusion}
\label{sec:disc}

The limitations of Markovianity, resulting from the absence of the {\sc qfdt}, have been analysed within the framework of two distinct models of damped harmonic oscillators, 
namely the \emph{friction} and \emph{cavity models}, specified by the Langevin equations~(\ref{eq:dynBMa},~\ref{eq:dynBMb}) and~(\ref{eq:dynCava},~\ref{eq:dynCavb}), respectively. 
These models have been studied countless times in the literature, see e.g. 
\cite{CaldeiraLegget1983,lindblad1976brownian,lindblad1976generators,bedeaux2001mesoscopic,bedeaux2002non,kohen1997,Rama2009,Grabert1984,HuPazZhang1992,oliveira2006quantum,isar1994open,isar2003lindblad,isar2006damped} or the textbooks including \cite{breuer2002theory,cohen2001,gardiner2004quantum,Weiss2022,carmichael1999statistical,Caldeira2014}.
Formally, the models differ in the explicit form of the non-stochastic dissipative terms but do rely on the same
physical criteria (A-D) on observables, used to fix the noise correlators. This approach allows us to circumvent an explicit treatment of initial system-bath correlations, 
and the many controversies on this which arose in the literature. 

In this work, we have demonstrated that the Markovian description gives rise to distinct inconsistencies in both models, as summarized in Tab.~\ref{fig:tab}.
We stress that the inconsistencies already arise in attempts of a Markovian ``quantum'' dynamics of the supposedly `simple' harmonic oscillator.
While the classical $\hbar\to 0$ limit of the friction model reduces to a Brownian particle 
in a harmonic potential (in agreement with Ehrenfest's theorem), 
no known classical system emerges for the cavity model, where Ehrenfest's theorem does not hold. 
{}From these Langevin equations, the equivalent master equations for the density operators are derived, by requiring identical equations of motion for one-point and two-point functions. 
While the cavity model is mapped to a Lindblad equation that ensures complete positivity and the semigroup property, the friction model yields a dynamical propagator that is not {\sc cp}. 
Reformulating the master equations as Fokker-Planck equations using the Wigner function formalism reveals that, in the absence of an external field, 
both models evolve toward the same thermal state. However, in the presence of a non-vanishing magnetic field, only the friction model ensures proper thermalisation.
Moreover, while the {\sc cp} cavity model violates spatial translation-invariance~\cite{kohen1997}, 
the friction model preserves both translation-invariance and quantum-classical correspondence, even though, for sufficiently squeezed initial states, 
its dynamical propagator can still generate non-positive states~\cite{talkner1986}.  
Explicit solutions of the equations of motion clearly illustrate the differences in the relaxation dynamics of the friction and cavity models: 
the friction model exhibits both \emph{under-damped} and \emph{over-damped} regimes, whereas the cavity model shows only an \emph{under-damped} regime. 
Nevertheless, in the weak damping limit ($\gamma \ll 1$), both models converge to the same phenomenology for time-dependent expectation values, as expected.
Finally, we proved that, when recast as mean-field approximation of a magnetic many-body system, 
the models are distinguished through a determination of the stationary state in an external field and the ensuing phase transition, see Fig.~\ref{fig:fig1}.

{}From a formal point of view, the failure of any model to obey all physical requirements (see Tab.~\ref{fig:tab}) 
should stem from the Markovian approximation.\footnote{Indeed, for the harmonic oscillator with {\em non}-Markovian quantum noise, all entries in Tab.~\ref{fig:tab} can be checked 
\cite{ford1965,ford1988,Weiss2022,araujo2019axiomatic,gardiner2004quantum,hanggi2005fundamental}.} 
Overall, there is no definitive criterion for favouring one model over another—or over alternative formulations with different dissipative terms, 
see~\cite{lindblad1976brownian,kohen1997}. The natural-looking requirement of complete positivity is not always convincing, 
as the cavity model neither satisfies Ehrenfest's theorem nor reduces, in the classical limit $\hbar \to 0$, to a classical Brownian particle in a harmonic potential.

{\em Non}-Markovian quantum dynamics can prevent the usually expected \cite{LandauLifschitz1979,PathriaBeale2011} relaxation towards the Gibbs-Boltzmann state, which
are nowadays described by the {\sc hmf} \cite{Miller2018,Trushechkin2022,Weiss2022}. While the
(Markovian) cavity model model for $B\ne 0$ does not appear to fit into this context,\footnote{This suggests that the innocent-looking addition of a magnetic field $B\ne 0$ in the cavity model leads to more deep properties, whose elucidation is beyond the scope of the present work.} a recent explicit
example occurs in the quantum spherical model, which is a set of harmonic oscillators coupled via the
so-called spherical constraint \cite{henkel1984,nieuwenhuizen1995quantum,vojta1996quantum,oliveira2006quantum}. 
Quantum dynamics occurs at bath temperature $T=0$ and the exact solution of the {\em non}-Markovian quantum Langevin equations shows that the stationary long-time limit
is not the Gibbs-Boltzmann state of equilibrium quantum statistical mechanics \cite{wald2021non,henkel2023quantum}. 
Our work is also meant as a preparation to study the many-body dynamics, notably in the quantum spherical model, where the spherical constraint renders the dynamics non-trivial \cite{wald2016lindblad}. 

Finally, we have seen that the choice of the dynamics can lead to drastic differences in the phase diagramme of quantum many-body systems, see Fig.~\ref{fig:fig1}, 
which calls for care in the choice of the quantum dynamics. 
The study of relaxational dynamics in the Markovian approximation of many-body systems and the comparison with the results of full quantum dynamics 
(respecting the {\sc qfdt}), e.g. \cite{chandran2013,sieberer2015thermodynamic,barbier2019pre}, will be taken up elsewhere. 

Therefore, the detailed comparisons carried out in this study have allowed us to critically revisit several widely-accepted statements, and we have given
an explicit and physically well-founded counter-example to the requirement of complete positivity in quantum master equations. 
Still, the inherent problems of the Markovian approximation imply that for a physically
consistent description a {\em non}-Markovian approach is needed, in agreement with the {\sc qfdt} and the fundamentally {\em non}-Markovian nature of the dynamics of 
open quantum systems ~\cite{ford1965,ford1988,gardiner2004quantum,hanggi2005fundamental,sieberer2015thermodynamic,araujo2019axiomatic,Weiss2022,tupkary2022}.

~\\
\noindent {\bf Acknowledgements:} It is a pleasure to thank D. Karevski and G.T. Landi for useful discussions.  
This work was supported by ANR-PRME UNIOPEN (ANR-22-CE30-0004-01). 
M.C. thanks the LUE for funding a  travel grant (Aide \`a la mobilit\'e internationale DrEAM). 
M.C. acknowledges the support of P1-0044 program of the Slovenian Research Agency and ERC StG 2022 project DrumS, Grant Agreement 101077265.
 
\newpage 

\appendix
\section{Noise correlators for the cavity model}
\label{appendixA}
We outline the solution of Eqs.~(\ref{eq:2.10a},~\ref{eq:2.10b}) and, subsequently, the derivation of the 
second moments~(\ref{eq:dynCavc},~\ref{eq:dynCavd},~\ref{eq:dynCave}) from the conditions (A-D). 
The lines of calculation follow closely Ref.~\cite{araujo2019axiomatic}. 

From the homogeneous part of Eqs.~(\ref{eq:2.10a},~\ref{eq:2.10b}), we find the relaxation rates $\Lambda_{\pm}=\frac{\gamma}{2} \pm \II \omega$, which suggests the ansatz
\begin{equation}
\hat{x}(t) = \hat{x}_{+}(t)\, \e^{-\Lambda_+ t} + \hat{x}_{-}(t)\, \e^{-\Lambda_- t}\,,\qquad
\hat{p}(t) = \hat{p}_{+}(t)\, \e^{-\Lambda_+ t} + \hat{p}_{-}(t)\, \e^{-\Lambda_- t}\,.
\end{equation}
Two of the four amplitudes in the above ansatz may still be chosen freely. 
Our choice is $\hat{p}_{\pm}(t) = - \frac{1}{g}\bigl( \Lambda_{\pm} - \frac{\gamma}{2}\bigr) \hat{x}_{\pm}(t) = \mp\frac{\II \omega}{g} \hat{x}_{\pm}(t)$. 
Insertion of the ansatz into~(\ref{eq:2.10a},~\ref{eq:2.10b}) leads to 
\begin{equation}
\partial_t \hat{x}_{+}(t) \, \e^{-\Lambda_+ t} + \partial_t \hat{x}_{+}(t) \, \e^{-\Lambda_+ t} = \hat{\eta}^{(x)}_c(t)\,,\qquad
-\frac{\II \omega}{g} \partial_t \hat{x}_{+}(t) \, \e^{-\Lambda_+ t} + \frac{\II \omega}{g} \partial_t \hat{x}_{+}(t) \, \e^{-\Lambda_+ t} = B + \hat{\eta}_c^{(p)}(t)\,.
\end{equation}
From these, we directly have
\begin{equation}
    \partial_t \hat{x}_{\pm}(t) = \demi \left( \hat{\eta}_c^{(x)}(t) \pm \frac{\II g}{\omega} \left( \hat{\eta}_c^{(p)} + B \right) \right) \,e^{\Lambda_{\pm} t}\,,
\end{equation}
which upon integration produces the solutions~(\ref{eq:2.11a},~\ref{eq:2.11b}).

The second moments~(\ref{eq:dynCavc},~\ref{eq:dynCavd},~\ref{eq:dynCave}) of the noises are found as follows. 
For the averaged noise commutator, we make the ansatz $\bigl\langle \left[ \hat{\eta}_c^{(x)}(t), \hat{\eta}_c^{(p)}(t') \right] \bigr\rangle =\kappa \delta(t-t')$. 
Then, the commutator $\bigl[ \hat{x}(t),\hat{p}(t')\bigr]=2C_{-}^{(x,p)}(t,t')$, computed using~(\ref{eq:2.11a},~\ref{eq:2.11b}),
\begin{equation}
2C_{-}^{(x,p)}(t,t')= \left( \frac{\II \omega}{g} \bigl\langle \left[ \hat{x}_{+}(0), \hat{x}_{-}(0) \right] \bigr\rangle - \frac{\kappa}{2\gamma} \right)
\left( \e^{-\Lambda_+ t - \Lambda_- t'} + \e^{-\Lambda_- t - \Lambda_+ t'} \right)
+ \frac{\kappa}{2\gamma} \left( \e^{-\Lambda_+ |t-t'|} + \e^{-\Lambda_- |t-t'|} \right)\,,
\end{equation}
contains a stationary part depending only on $|t-t'|$ and a rapidly decaying transient part. 
In order to achieve the equal-time canonical position-momentum commutator (A), the transient term must vanish which is achieved through the choice 
\begin{equation} \label{eq:A.6}
\bigl\langle \left[ \hat{x}_{+}(0), \hat{x}_{-}(0) \right] \bigr\rangle = \frac{1}{\II} \frac{\kappa g}{2\gamma \omega}\,.
\end{equation}
Then the averaged equal-time commutator $\bigl\langle \left[ \hat{x}(t), \hat{p}(t) \right] \bigr\rangle = \frac{\kappa}{\gamma}$ 
is fixed to the canonical value (A) by choosing 
\begin{equation}\label{eq:A.7}
    \kappa\stackrel{!}{=}\II \hbar\gamma\,,
\end{equation}
which gives the noise correlator~(\ref{eq:dynCavc}). Next, we check the Kubo formula (B). On one side, with the definitions (\ref{eq:reponse}), we obtain 
\begin{equation}
\partial_t R^{(x)}(t,s) = g R^{(p)}(t,s) - \frac{\gamma}{2} R^{(x)}(t,s)\,,\qquad
\partial_t R^{(p)}(t,s) = -\frac{\omega^2}{g} R^{(x)}(t,s) - \frac{\gamma}{2} R^{(p)}(t,s) + \delta(t-s)\,.
\end{equation}
Since these only depend on the time difference $\tau=t-s$, we define $\mathscr{R}(\tau) := R^{(x)}(\tau+s,s)=R^{(x)}(t,s)$, which obeys the second-order equation
\begin{equation}
\partial_{\tau}^2 \mathscr{R}(\tau) + \gamma\partial_{\tau} \mathscr{R}(\tau) + \left( \omega^2 + \frac{\gamma^2}{4} \right) \mathscr{R}(\tau) = g \delta(\tau)\,.
\end{equation}
Using the Fourier transform $\wht{\mathscr{R}}(\nu) = \frac{1}{\sqrt{2\pi\,}} \int_{\mathbb{R}} \!\D \tau\: \e^{-\II\nu\tau} \mathscr{R}(\tau)$, we find
\begin{equation} \label{eq:A.9}
\wht{\mathscr{R}}(\nu)= - \frac{g}{\sqrt{2\pi\,}} \frac{1}{\nu^2 - \II\gamma \nu - \omega^2 - \gamma^2/4}\,.
\end{equation}
whose p\^oles $\nu_{\pm}=\II\frac{\gamma}{2} \pm \omega$ are in the upper complex $\nu$-half-plane. 
Consequently, inverting (\ref{eq:A.9}) via the residue theorem, the response of the position $x$ reads
\begin{equation}
\mathscr{R}(\tau) = R^{(x)}(\tau+s,s) = \frac{g}{\omega}\, \e^{-\gamma\tau/2} \sin \bigl(\omega \tau\bigr)\: \Theta\bigl(\tau\bigr)\,,
\end{equation}
where the Heaviside function $\Theta(\tau)$ expresses the expected causality. On the other hand, Eq.~\eqref{eq:2.11a} implies
\begin{equation}
C_{-}^{(x,x)}(t,t') = \demi\left( \bigl\langle \bigl[ \hat{x}_{+}(0), \hat{x}_{-}(0) \bigr] \bigr\rangle- \frac{\hbar g}{2\omega} \right) 
\left( \e^{-\Lambda_{+}t - \Lambda_{-} t'} - \e^{-\Lambda_{+}t' - \Lambda_{-} t} \right) 
+ \frac{\hbar}{2\II} \frac{g}{\omega} \e^{-\gamma |t-t'|/2} \sin\bigl( \omega |t-t'|\bigr)\,.
\end{equation}
However, from Eqs.~(\ref{eq:A.6},~\ref{eq:A.7}) we get $\bigl\langle \bigl[ \hat{x}_{+}(0), \hat{x}_{-}(0) \bigr] \bigr\rangle=\frac{\hbar g}{2\omega}$, hence
the Kubo formula $R^{(x)}(\tau+s,s) = \frac{2\II}{\hbar} C_{-}^{(x,x)}(\tau+s,s) \Theta(\tau)$ is satisfied. 

Now, we impose the virial theorem (C).
Following~\cite{araujo2019axiomatic}, we make the ansatz
\begin{equation} \label{eq:A.12}
\frac{1}{2}\bigl\langle \left\{ \hat{\eta}_c^{(p)}(t), \hat{\eta}_c^{(p)}(t') \right\} \bigr\rangle = \alpha \,\delta(t-t')\,,\quad
\frac{1}{2}\bigl\langle \left\{ \hat{\eta}_c^{(x)}(t), \hat{\eta}_c^{(x)}(t') \right\} \bigr\rangle = \beta\,\delta(t-t')\,,\quad
\bigl\langle \left\{ \hat{\eta}_c^{(x)}(t), \hat{\eta}_c^{(p)}(t') \right\} \bigr\rangle = 0\,.
\end{equation}
To find the constants $\alpha,\beta$, first we calculate 
\begin{align}
\hspace{-0.5cm}C_{+,{\rm c}}^{(x,x)}(t,t') =& \left( \bigl\langle \hat{x}_{+}(0)^2\bigr\rangle - \frac{\beta-\alpha g^2/\omega^2}{8 \Lambda_+}\right) \e^{-\Lambda_+ (t+t')} 
+ \left( \bigl\langle \hat{x}_{-}(0)^2\bigr\rangle - \frac{\beta-\alpha g^2/\omega^2}{8 \Lambda_-}\right) \e^{-\Lambda_- (t+t')} \nonumber \\
 & \hspace{-0.5cm}+ \left( \frac{1}{2}\bigl\langle \bigl\{ \hat{x}_{+}(0), \hat{x}_{-}(0)\bigr\}\bigr\rangle - \frac{\beta+\alpha g^2/\omega^2}{4( \Lambda_++\Lambda_{-})}\right) 
\left( \e^{-\Lambda_+  t - \Lambda_{-} t'} +\e^{-\Lambda_+  t' - \Lambda_{-} t}\right)  \nonumber \\
 & \hspace{-0.5cm}+ \frac{1}{4} \left( \frac{\beta - \alpha g^2/\omega^2}{2\Lambda_{+}} 
    + \frac{\beta + \alpha g^2/\omega^2}{\Lambda_{+} +\Lambda_{-}} \right) \e^{-\Lambda_{+} |t-t'|} 
    + \frac{1}{4} \left( \frac{\beta - \alpha g^2/\omega^2}{2\Lambda_{-}} 
    + \frac{\beta + \alpha g^2/\omega^2}{\Lambda_{+} +\Lambda_{-}} \right) \e^{-\Lambda_{-} |t-t'|}\,.
\label{eq:A.13}
\end{align}
The terms in the first two lines are rapidly decaying transient terms whereas the stationary terms are collected in the third line 
and the non-connected terms $\sim B^2$ cancelled. 
Second, we similarly obtain
\begin{align}
\hspace{-0.5cm}C_{+,{\rm c}}^{(p,p)}(t,t') =& 
\frac{\omega^2}{g^2}\left( - \bigl\langle \hat{x}_{+}(0)^2\bigr\rangle + \frac{\beta-\alpha g^2/\omega^2}{8 \Lambda_+}\right) \e^{-\Lambda_+ (t+t')} 
+ \frac{\omega^2}{g^2}\left( - \bigl\langle \hat{x}_{-}(0)^2\bigr\rangle + \frac{\beta-\alpha g^2/\omega^2}{8 \Lambda_-}\right) \e^{-\Lambda_- (t+t')} \nonumber \\
 & \hspace{-2cm}+ \frac{\omega^2}{g^2}\left( \frac{1}{2}\bigl\langle \bigl\{ \hat{x}_{+}(0), \hat{x}_{-}(0)\bigr\}\bigr\rangle 
    - \frac{\beta+\alpha g^2/\omega^2}{4( \Lambda_+ +\Lambda_{-})}\right) 
\left( \e^{-\Lambda_+  t - \Lambda_{-} t'} +\e^{-\Lambda_+  t' - \Lambda_{-} t}\right)  \nonumber \\
 & \hspace{-2cm}+ \frac{\omega^2}{4g^2}\left( -\frac{\beta - \alpha g^2/\omega^2}{2\Lambda_{+}} + \frac{\beta + \alpha g^2/\omega^2}{\Lambda_{+} +\Lambda_{-}} \right) 
    \e^{-\Lambda_{+} |t-t'|} 
+ \frac{\omega^2}{4g^2} \left( -\frac{\beta - \alpha g^2/\omega^2}{2\Lambda_{-}} + \frac{\beta + \alpha g^2/\omega^2}{\Lambda_{+} +\Lambda_{-}} \right) 
   \e^{-\Lambda_{-} |t-t'|} \,,
\label{eq:A.14}
\end{align}
with transient terms in the first two lines and the stationary contributions in the last line. Comparing (\ref{eq:A.13},\ref{eq:A.14}), 
\begin{equation}
\frac{1}{4} \left( \frac{\beta - \alpha g^2/\omega^2}{2\Lambda_{\pm}}  + \frac{\beta + \alpha g^2/\omega^2}{\Lambda_{+} +\Lambda_{-}} \right) \stackrel{!}{=}
\frac{1}{4} \left( -\frac{\beta - \alpha g^2/\omega^2}{2\Lambda_{\pm}} + \frac{\beta + \alpha g^2/\omega^2}{\Lambda_{+} +\Lambda_{-}} \right)\,,
\end{equation}
which leads to 
\begin{equation} \label{gl:equi}
\alpha = \frac{\omega^2}{g^2} \beta\,.
\end{equation}
Finally, the condition (D) specifies the position (connected) auto-correlator. Using the stationary part of Eq.~(\ref{eq:A.13}), we obtain 
\begin{equation}
C_{+,{\rm c, stat}}^{(x,x)} = \frac{\beta}{2\gamma} \stackrel{!}{=} \frac{\hbar g}{2\omega} \coth\left( \frac{\hbar\omega}{2 T}\right)\,,
\end{equation}
which implies 
\begin{equation}\label{beta_deriv}
\beta=\frac{\hbar \gamma g}{\omega}  \coth\left( \frac{\hbar\omega}{2 T}\right)\,.
\end{equation}
Insert Eqs.~(\ref{gl:equi},~\ref{beta_deriv}) into the ansatz~(\ref{eq:A.12}) to reproduce the noise correlators~(\ref{eq:dynCavc},~\ref{eq:dynCavd},~\ref{eq:dynCave}) in the main text. 

\section{Phase-space formulation of quantum mechanics}
\label{appendixB}
The phase-space formulation of quantum mechanics, introduced by Wigner \cite{wigner1932}, 
is attractive because of its similarity to classical Hamiltonian dynamics. 
It is based on mapping quantum states to quasi-probability distribution functions and the Weyl transform, 
which maps quantum operators into real-valued functions defined on 
phase-space $(x,p)$. This facilitates connections to classical mechanics and semi-classical limits \cite{moyal1949quantum,hillery1984,lee1995,schleich2001quantum,case2008}. 

Quantum observables are represented by Hermitian operators acting on the Hilbert space of system states, 
such that the outcome of a measurement must be an eigenvalue of the
observable. Quantisation establishes a map between classical variables and quantum operators, 
such that the canonical commutation relation $[\hat{x},\hat{p}]=\II\hbar$ between
position $\hat{x}$ and momentum $\hat{p}$ holds true. However, for non-linear observables in position and momentum, the operator ordering is ambiguous. 
Therefore, one must define a one-to-one map  $\Phi[\hat{f}]=f(x,p)$ between the Hilbert space operator $\hat{f}$ and its analytic phase-space representation $f(x,p)$. 
This map $\Phi$ uniquely defines a non-commutative product $\star$ between the phase-space representations $f(x,p)$, $g(x,p)$ of two Hilbert space operators 
$\hat{f}$, $\hat{g}$ such that
\begin{equation}\label{product_phase_space_functions}
    f(x,p)\star g(x,p)=\Phi[\hat{f}\hat{g}]\,,
\end{equation}
is the phase-space representation of the product Hilbert operator $\hat{f}\hat{g}$. 
Observe that the product $\star$ must be consistent with the commutation relation $[\hat{x},\hat{p}]=\II\hbar$ and, in the limit $\hbar\to 0$, 
it should reduce to the standard commutative product. 
However, the product $\star$ depends on the choice of $\Phi$, which allows for multiple possible quantization schemes. The {\em Weyl quantisation} is one way to carry out this programme. 

For any operator $\hat{f}$ acting on the Hilbert space, the {\em Weyl transform} is the functional $\Phi[\hat{f}]:=f(x,p)$ such that
\begin{equation}\label{Weyl_trnasform}
    \Phi[\hat{f}]=\int_{\mathbb{R}} e^{\frac{\rm i}{\hbar}pw}
    \hspace{0.1cm}\langle x-w/2\hspace{0.1cm}|\hspace{0.1cm}\hat{f}\hspace{0.1cm}|\hspace{0.1cm}x+w/2\rangle\hspace{0.1cm} \D w\,,
\end{equation}
using the eigenstates $|x\rangle$ of the position operator $\hat{x}\,|x\rangle=x\,|x\rangle$. 
The phase-space product $\star$ associated to the Weyl quantisation is called {\em Moyal product} \cite{moyal1949quantum} 
and satisfies Eq.~\eqref{product_phase_space_functions}. Let $f(x,p)$ and $g(x,p)$ 
be the phase-space representations of the operators $\hat{f}$ and $\hat{g}$. The product $\hat{f}\hat{g}$ maps onto 
\begin{equation}\label{MOYAL_product}
    \hat{f}\hat{g}\hspace{0.4cm}\longrightarrow \hspace{0.4cm}f(x,p)\star g(x,p) 
    := f(x,p)\exp\bigg(\frac{\II\hbar}{2}\Big(\overset{\leftarrow}{\partial}_x\overset{\rightarrow}{\partial}_p
      -\overset{\leftarrow}{\partial}_p\overset{\rightarrow}{\partial}_x \Big)\bigg)\hspace{0.1cm}g(x,p)\,,
\end{equation}
where the arrows indicate the direction for the differentiation. The commutator and the anti-commutator, respectively, are mapped onto 
\begin{align}
\left[\hat{f},\hat{g}\right]\hspace{0.4cm}&\longrightarrow  
      \hspace{0.4cm}2{\rm i} f(x,p)\sin\bigg(\frac{\hbar}{2}\Big(\overset{\leftarrow}{\partial}_x\overset{\rightarrow}{\partial}_p
      -\overset{\leftarrow}{\partial}_p\overset{\rightarrow}{\partial}_x \Big)\bigg)\,g(x,p)\,,\\
\bigl\{\hat{f},\hat{g}\bigl\}\hspace{0.4cm}&\longrightarrow
      \hspace{0.4cm}2 f(x,p)\cos\bigg(\frac{\hbar}{2}\Big(\overset{\leftarrow}{\partial}_x\overset{\rightarrow}{\partial}_p
      -\overset{\leftarrow}{\partial}_p\overset{\rightarrow}{\partial}_x \Big)\bigg)\,g(x,p)\,.
\end{align}

\section{Wigner function dynamics}
\label{appendixC}
Starting from the Weyl transform \eqref{Weyl_trnasform}, one defines the {\em Wigner function} 
\begin{equation}\label{Wigner_function}
    \mathcal{W}(x,p):=\frac{1}{2\pi\hbar}\int_{\mathbb{R}} 
    e^{\frac{\rm i}{\hbar}pw}\hspace{0.1cm}\langle x-w/2\hspace{0.1cm}|\hspace{0.1cm}\hat{\rho}\hspace{0.1cm}|\hspace{0.1cm}x+w/2\rangle\hspace{0.1cm} \D w\,,
\end{equation}
where $\hat{\rho}$ is the density operator of the system  (pure or mixed), assumed to be normalised. 
$\mathcal{W}(x,p)$ is the phase-space picture of the density matrix and fully characterises the system. The Wigner function has the proprieties \cite{case2008}:  
(i) $\mathcal{W}(x,p)$ is a real function with possibly negative values and $\bigl| \mathcal{W}(x,p)\bigr|\leq \bigl(\pi \hbar\bigr)^{-1}$; 
(ii) it is normalised $\iint \!\D x\D p\: \mathcal{W}(x,p)  =1$; 
(iii) the marginals of the Wigner function co\"{\i}ncide with the probability densities in position and momentum space; 
(iv) a phase-space representation $O(x,p)$ of the quantum observable $\hat{O}$ has the average value 
$\langle\hat{O}\rangle=\tr(\hat{O}\hat{\rho})= \iint \!\D x\D p\: O(x,p)\mathcal{W}(x,p)$. 
These features make $\mathcal{W}(x,p)$ a {\em quasi-probability distribution}. 
Tab.~\ref{tab1} lists common quantum operators with their respective Weyl transforms.

\begin{table}[!t]
    \centering
\begin{tabular}{ccc}
\begin{tabular}[t]{|c|c|} \hline
Quantum operator & Weyl transform\\ \hline
$\hat \rho$                                        & $\mathcal{W}(x,p)$ \\ \hline
$\hat H$                                           & $\frac{g}{2}p^2+\frac{\omega^2}{2g}x^2 - B x$ \\ \hline
$\hat H_f$                                         & $\frac{g}{2}p^2+\frac{\omega^2}{2g}x^2 - B x +\frac{\gamma}{2}xp$ \\ \hline 
$[\hat H,\hat \rho]$   & ${\rm i}\hbar\bigl[\bigl(\frac{\omega^2}{g}x-B\bigl)\partial_p\mathcal{W}-gp\partial_x\mathcal{W}\bigl]$ \\ \hline
$[\hat H_f,\hat \rho]$ & ${\rm i}\hbar\bigl[\bigl(\frac{\omega^2}{g}x-B+\frac{\gamma}{2}p\bigl)\partial_p\mathcal{W}-\bigl(gp+\frac{\gamma}{2}x\bigl)\partial_x\mathcal{W}\bigl]$ \\ \hline
$\hat p \hat \rho \hat x$    & $xp\mathcal{W}-\frac{{\rm i}\hbar}{2}\bigl(x\partial_x\mathcal{W}+p\partial_p\mathcal{W}+\mathcal{W}\bigl)
            -\bigl(\frac{\hbar}{2}\bigl)^2\partial_x\partial_p\mathcal{W}$ \\ \hline
$\hat x \hat \rho \hat p$    & $xp\mathcal{W}+\frac{{\rm i}\hbar}{2}\bigl(x\partial_x\mathcal{W}+p\partial_p\mathcal{W}+\mathcal{W}\bigl)
            -\bigl(\frac{\hbar}{2}\bigl)^2\partial_x\partial_p\mathcal{W}$ \\ \hline
$D_{\rho}(\hat x,\hat x)$ & $\frac{\hbar^2}{2}\partial_p^2\mathcal{W}$\\ \hline 
$D_{\rho}(\hat p,\hat p)$ & $\frac{\hbar^2}{2}\partial_x^2\mathcal{W}$\\ \hline 
\end{tabular} \tabularnewline
\end{tabular}
\caption{Some quantum operators with their respective Weyl transforms. For simplicity, in this table we used units $2\pi\hbar=1$. \label{tab1} }
\label{table}
\end{table}

The equations of motion of the Wigner function of the friction and cavity models  follow from the master equations~\eqref{eq:3.13} and~\eqref{eq:cavity}. 
Using the canonical transformation~\eqref{eq:aa} in Eqs.~(\ref{eq:3.13},~\ref{eq:cavity}), 
the dissipators of the friction and cavity models read
\begin{equation}\label{frict-xp}
\mathcal{D}(\hat{\rho}_f)= (2 n_\omega + 1) \frac{\omega\gamma}{\hbar g}\,D_{\rho_f}(\hat x,\hat x)
+\frac{\gamma}{2}\hat\rho_f- \frac{{\rm i}\gamma}{2\hbar}(\hat x \hat \rho_f\hat p - \hat p \hat \rho_f \hat x)\,,    
\end{equation}
\begin{equation}\label{cav-xp}
\mathcal{D}(\hat{\rho}_c)= (2 n_\omega + 1) \frac{\omega\gamma}{2\hbar g}\,D_{\rho_c}(\hat x,\hat x) 
+ (2 n_\omega + 1) \frac{g\gamma}{2\hbar \omega}\,D_{\rho_c}(\hat p,\hat p)
+\frac{\gamma}{2}\hat\rho_c- \frac{{\rm i}\gamma}{2\hbar}(\hat x \hat \rho_c\hat p - \hat p \hat \rho_c \hat x)\,.
\end{equation}
Using the Weyl transforms in Tab.~\ref{table} and Eqs.~(\ref{frict-xp},\ref{cav-xp}), 
the master equations~(\ref{eq:3.13},~\ref{eq:cavity}) are finally mapped into
\begin{equation}\label{aa}
    \partial_t \mathcal{W}_f=(2 n_\omega + 1) \frac{\hbar\gamma}{2}\frac{\omega}{g}\,\partial_p^2\mathcal{W}_f
            +\bigg(\frac{\omega^2}{g}x+\gamma p-B\bigg)\partial_p\mathcal{W}_f-gp\partial_x\mathcal{W}_f+\gamma\,\mathcal{W}_f\,,
\end{equation}
\begin{equation}\label{ab}
    \partial_t \mathcal{W}_c=\frac{\hbar\gamma}{4 }(2 n_\omega + 1)\bigg[\frac{g}{\omega}\partial_x^2\mathcal{W}_c
           +\frac{\omega}{g}\partial_p^2\mathcal{W}_c\bigg]+\bigg(\frac{\omega^2}{g}x+\frac{\gamma}{2} p-B\bigg)\partial_p\mathcal{W}_c
           +\bigg(\frac{\gamma}{2}x-gp\bigg)\partial_x\mathcal{W}_c+\gamma\,\mathcal{W}_c\,.
\end{equation}
Observe that, in the limit $\hbar\omega/T\ll 1$, Eq.~(\ref{aa}) reduces to the form quoted in \cite[Eq. (6.194)]{Weiss2022}. 
With some simple algebra, Eqs.~(\ref{aa}) and (\ref{ab}) can be rewritten in a more compact form as follows, 
\begin{equation}\label{F:W-evo}
        \partial_t \mathcal{W}_\alpha(t,x,p) =\Vec{\nabla} \bigcdot \left[\frac{\Vec{D}_\alpha}{2}\Vec{\nabla}\mathcal{W}_\alpha(t,x,p)+\Vec{\Theta}_\alpha
    \begin{pmatrix}x \\ p \end{pmatrix}\mathcal{W}_\alpha(t,x,p)
    -\Vec{\Theta}_\alpha\Vec{\mu}_\alpha \mathcal{W}_\alpha(t,x,p)\right]\,,
\end{equation}
where $\alpha=f,c$ for the friction ($f$) and cavity ($c$) models,  
\begin{equation}\label{F:notations}
\Vec {D}_{f}= \begin{pmatrix}0 &0 \\
    0 & \frac{\omega\hbar\gamma}{g}\Big(2n_\omega+1\Big) 
    \end{pmatrix}\,,\quad \Vec{\Theta}_{f} = \begin{pmatrix}0 &-g \\
    {\omega^2}/{g} & \gamma
    \end{pmatrix}\,,\quad\Vec{\mu}_{f}=B\begin{pmatrix}{g}/{\omega^2}\\0
    \end{pmatrix}\,,
\end{equation}
\begin{equation}
    \Vec{D}_{c}= \begin{pmatrix}\frac{g\hbar\gamma}{2\omega}\Big(2n_\omega+1\Big)  &0 \\
    0 & \frac{\omega\hbar\gamma}{2g}\Big(2n_\omega+1\Big) 
    \end{pmatrix}\,,\quad \Vec{\Theta}_{c} = \begin{pmatrix}\gamma/2 &-g \\
    {\omega^2}/{g} & \gamma/2
    \end{pmatrix}\,,\quad\Vec{\mu}_{c}=B\bigl({\omega^2+\gamma^2/4}\bigr)^{-1}\begin{pmatrix}g\\ {\gamma}/2\end{pmatrix}\,,
\end{equation}
and $n_{\omega}=\bigl(\e^{\hbar\omega/T}-1\bigr)^{-1}$. Eq.~\eqref{F:W-evo} is a Fokker-Plank equation, which is solved by standard analytical techniques. 
To begin, we define the Fourier transform with respect to the reciprocal space variable $\Vec{k}$ as
\begin{equation}
\wit{\mathcal{W}}_\alpha(t,\vek{k}):=\int_{\mathbb{R}^2} \!\D\Vec{r}\hspace{0.1cm}\mathcal{W}_\alpha(t,\Vec{r})\, \e^{-\II \Vec{r}\cdot\Vec{k}/\hbar}, \qquad 
    \Vec{r}=\begin{pmatrix}x \\ p \end{pmatrix}\,, 
    \quad\Vec{k}=\begin{pmatrix}\tilde p \\ \tilde x \end{pmatrix}\,.    
\end{equation}
A standard calculation gives the time-dependence of the transformed Wigner function~\cite{vatiwutipong2019alternative},
\begin{align}
    \label{F:eq:Wtilde}
    \wit{\mathcal{W}}_\alpha(t,\Vec{k}) =&\; \wit{\mathcal{W}}_\alpha\Big(0,\e^{-\Vec{\Theta}_\alpha^T t}\Vec{k}\Big)\: 
    \exp{-\frac{\II}{\hbar}\Vec{k}^T\Big(\mathbf{1}-\e^{-\Vec{\Theta}_\alpha t}\Big)\Vec{\mu}_\alpha- \frac{1}{2\hbar^2}\Vec{k}^T\Vec{\Sigma}_\alpha(t)\Vec{k}}\,.\\
   \label{F:eq:sigma}
   \Vec{\Sigma}_\alpha(t) :=&\; \int_0^t \!\D s\hspace{0.1cm}\e^{-s\Vec{\Theta}_\alpha}\Vec{D}_\alpha\,\e^{-s\Vec{\Theta}_\alpha^T} = \Vec{\Sigma}_\alpha(t)^T   \,. 
\end{align}
We observe that $\wit{\mathcal{W}}_\alpha(t,\Vec{k})$ is given by the product between an exponential function and 
$\wit{\mathcal{W}}_\alpha\Big(0,\e^{-\Vec{\Theta}_\alpha^T t}\Vec{k}\Big)$. 
The inverse Fourier transforms of 
$\exp{-\frac{\II}{\hbar}\Vec{k}^T\Big(\mathbf{1}-\e^{-\Vec{\Theta}_\alpha t}\Big)\Vec{\mu}_\alpha- \frac{1}{2\hbar^2}\Vec{k}^T\Vec{\Sigma}_\alpha(t)\Vec{k}}$ 
and $\wit{\mathcal{W}}_\alpha\Big(0,\e^{-\Vec{\Theta}_\alpha^T t}\Vec{k}\Big)$ are \begin{align}
    f_\alpha(t,\Vec{r}):=&\;\int_{\mathbb{R}^2} \frac{\D\Vec{k}}{(2\pi\hbar)^2}
    \exp\bigg\{-\frac{\II}{\hbar}\Vec{k}^T\Big(\mathbf{1}-\e^{-\Vec{\Theta}_\alpha t}\Big)\Vec{\mu}_\alpha
               - \frac{1}{2\hbar^2}\Vec{k}^T\Vec{\Sigma}_\alpha(t)\Vec{k}\bigg\}\e^{\II\Vec{k}\cdot\Vec{r}/\hbar}
    \nonumber\\
    =&\; \frac{1}{2\pi\sqrt{\det\Vec{\Sigma}_\alpha(t)}}\exp{-\frac{1}{2}\Big(\Vec{r}-\Vec{m}_\alpha(t)\Big)^T\Vec{\Sigma}_\alpha^{-1}(t)
       \Big(\Vec{r}-\Vec{m}_\alpha(t)\Big)}\label{eq:f_t}\,,\\
    \mathcal{W}'_\alpha(t,\Vec{r}) :=&\; \int_{\mathbb{R}^2} \frac{\D\Vec{k}}{(2\pi\hbar)^2}
    \wit{\mathcal{W}}_\alpha\Big(0,\e^{-\Vec{\Theta}_\alpha^T t}\Vec{k}\Big)\,\e^{\II\Vec{k}\cdot\Vec{r}/\hbar} 
    \:=\: \e^{\tr(\Vec{\Theta}_\alpha)t}\,\mathcal{W}_\alpha(0,\e^{\Vec{\Theta}_\alpha t}\Vec{r})\,,\label{eq:W'}
\end{align} 
with the initial distribution $\mathcal{W}_\alpha(0,\Vec{r})$ and $\Vec{m}_\alpha(t)=\Big(\mathbf{1}-\e^{-\Vec{\Theta}_\alpha t}\Big)\Vec{\mu}_\alpha$. 
Using the convolution theorem, the full dynamics is given by 
\begin{equation} \label{F:conv}
    \mathcal{W}_\alpha(t,\Vec{r})= \int_{\mathbb{R}^2} \!\D \Vec{r'}\: f_\alpha(t,\Vec{r}-\Vec{r'})\mathcal{W}'_\alpha(t,\Vec{r'})\,.
\end{equation}
We remark that the dynamics generated by Eq.~\eqref{F:W-evo} is Gaussian-preserving and this feature clearly emerges from Eq.~\eqref{F:conv}: 
$f_\alpha(t,\Vec{r})$ is a Gaussian function and the convolution of two Gaussians is Gaussian as well. 

From Eq.~\eqref{eq:f_t}, the evolution is governed by the matrix $\Vec{\Sigma}_\alpha(t)$, which we now evaluate. 
Using the Cayley–Hamilton theorem~\cite{Householder53,GlossarWiki:Fischer}, 
the $2\times 2$ exponential matrix $\e^{-\Vec{\Theta}_\alpha t}$ may be written as 
\begin{equation}
    \e^{-\Vec{\Theta}_\alpha t}=s^\alpha_0(t)\mathbf{1}+s^\alpha_1(t)\Vec{\Theta}_\alpha\,,
\end{equation}
with 
\begin{equation}
    s_0^\alpha(t)=\frac{\Lambda_+^{(\alpha)} \e^{-\Lambda_{-}^{(\alpha)}t}-\Lambda_{-}^{(\alpha)}\e^{-\Lambda_+^{(\alpha)}t}}{\Lambda_+^{(\alpha)}-\Lambda_{-}^{(\alpha)}}\,,\qquad
    s^\alpha_1(t)=\frac{\e^{-\Lambda_{-}^{(\alpha)}t}-\e^{-\Lambda_{+}^{(\alpha)}t}}{\Lambda_+^{(\alpha)}-\Lambda_{-}^{(\alpha)}}\,,
\end{equation}
and the ({\it a priori} complex and distinct) eigenvalues  $\Lambda_{\pm}^{(\alpha)}$ 
of $\Vec{\Theta}_\alpha$, see Sec.~\ref{sec:mod}.  
Eq.~\eqref{F:eq:sigma} reduces to 
\begin{equation}\label{sigma_t}
   \Vec{\Sigma}_\alpha(t)=\int_0^t \!\D t'\hspace{0.1cm}s^\alpha_0(t')^2\Vec{D}_\alpha+
   \int_0^t \!\D t'\hspace{0.1cm}s^\alpha_1(t')^2\Vec{\Theta}_\alpha \Vec{D}_\alpha\Vec{\Theta}_\alpha^T+
   \int_0^t \!\D t'\hspace{0.1cm}s^\alpha_0(t')s^\alpha_1(t')\Big(\Vec{\Theta}_\alpha \Vec{D}_\alpha+\Vec{D}_\alpha\Vec{\Theta}_\alpha^T\Big)\,,
\end{equation}
with the same long-time limit in both models 
\begin{equation} \label{F12}
\Vec{\Lambda}_T:=\lim_{t\to\infty}\Vec{\Sigma}_\alpha(t)=\int_0^\infty \!\D t\: \e^{-\Vec{\Theta}_\alpha t}\Vec{D}_\alpha\,\e^{-\Vec{\Theta}_\alpha^T t}
    =\frac{\hbar}{2}\begin{pmatrix} g/\omega & 0 \\
                     0         & \omega/g
    \end{pmatrix}\coth{\frac{\omega\hbar}{2T}}\,.
\end{equation}

\section{Dynamics of Gaussian states}
\label{appendixD}

As an example, we consider the system to be prepared in a Gaussian state, e.g. any Gibbs thermal state for $\hat{H}$, where 
\begin{equation} \label{F17} 
    \mathcal{W}_\alpha(0,\Vec{r})=\frac{1}{2\pi\sqrt{\det\Vec{\Gamma}\,}\,}\exp{-\frac{1}{2}(\Vec{r}-\Vec{\nu})^T\Vec{\Gamma}^{-1}(\Vec{r}-\Vec{\nu})}\,,
\end{equation}
for the initial matrix $\Vec{\Gamma}$ and the vector $\Vec{\nu}$. 
Our choice of Gaussian states is less restrictive as it might appear at first sight. 
Namely, wavelet theory shows that states represented as an infinite sum of shifted Gaussians are dense in the space of (square-)integrable functions 
\cite[sect. 6.6]{meyer1992wavelets},\cite{Mazya2007,calcaterra2008approximating}. 
Hence any solution can be obtained, to arbitrary precision, by summing over Gaussian contributions.
According to Eq.~(\ref{F:conv}), the full solution under the initial condition~(\ref{F17}) reads
\begin{equation} \label{F:Wigner-sol}
    \mathcal{W}_\alpha(t,\Vec{r})=\frac{1}{2\pi\sqrt{\det\Vec{\Omega}_\alpha(t)\,}\,}
    \exp{-\frac{1}{2}\Big(\Vec{r}-\Vec{M}_\alpha(t)\Big)^T\Vec{\Omega}_\alpha^{-1}(t)\Big(\Vec{r}-\Vec{M}_\alpha(t)\Big)}\,,
\end{equation}
where 
\begin{equation}\label{F:cap_omega}
\Vec{M}_\alpha(t)=\e^{-\Vec{\Theta}_\alpha t}\Vec{\nu}+\left( 1- e^{-\Vec{\Theta}_{\alpha}t} \right)\Vec{\mu}_\alpha\,,\qquad
\Vec{\Omega}_\alpha(t)=\e^{-\Vec{\Theta}_\alpha t}\Vec{\Gamma}\e^{-\Vec{\Theta}_\alpha^T t}+\Vec{\Sigma}_\alpha(t)\,.
\end{equation}
Therefore, the expectation values of observables are given by 
\begin{equation}
\int_{\mathbb{R}^2} \!\D\Vec{r}\: O(\Vec{r})\mathcal{W}_\alpha(t,\Vec{r})=  
\exp{\frac{1}{2}\sum_{i,j=1}^2(\Vec{\Omega}_\alpha)_{ij}\partial_{r_i}\partial_{r_j}}O\Big(\Vec{r}+\Vec{M}_\alpha(t)\Big)\Bigg|_{\Vec{r}=0}\,,   
\end{equation}
where $O(\Vec{r})$ is the Weyl transform of the observable $\hat{O}$.

We already mentioned that, in the friction model, strongly squeezed initial states can lead to non-positivity~\cite{talkner1986}. 
We now illustrate this through an analysis of positivity in the context of Gaussian states. 
Studying the conditions under which the density operator remains positive semi-definite amounts to analysing the evolution of the matrix $\Vec{\Omega}_\alpha(t)$, 
which must itself be positive in order to generate physically admissible density matrices. This, in turn, requires that the eigenvalues of 
$\Vec{\Omega}_\alpha(t)$ be strictly positive and consistent with the Heisenberg principle.
While the positivity of $\Vec{\Omega}_\alpha(t)$ 
is automatically guaranteed for the cavity model~\cite{lindblad1976generators}, the same does not hold for the friction model, as is well-known \cite{talkner1986}. 
On the other hand, since the long-time limit $\Vec{\Lambda}_T=\lim_{t\to\infty} \Vec{\Sigma}_{\alpha}(t)$ is positive definite, see (\ref{F12}), 
both models lead to positive semi-definite density matrices for large enough times and non-positivity can only arise at short and intermediate times in the friction model. 
Following~\cite{talkner1986}, and given the parameters $\omega$, $g$, $B$, $T$, and $\gamma$, we focus on the \emph{sufficient} initial-state conditions that ensure short-time positive dynamics.

\begin{figure}
    \centering
    \includegraphics[width=0.5\linewidth]{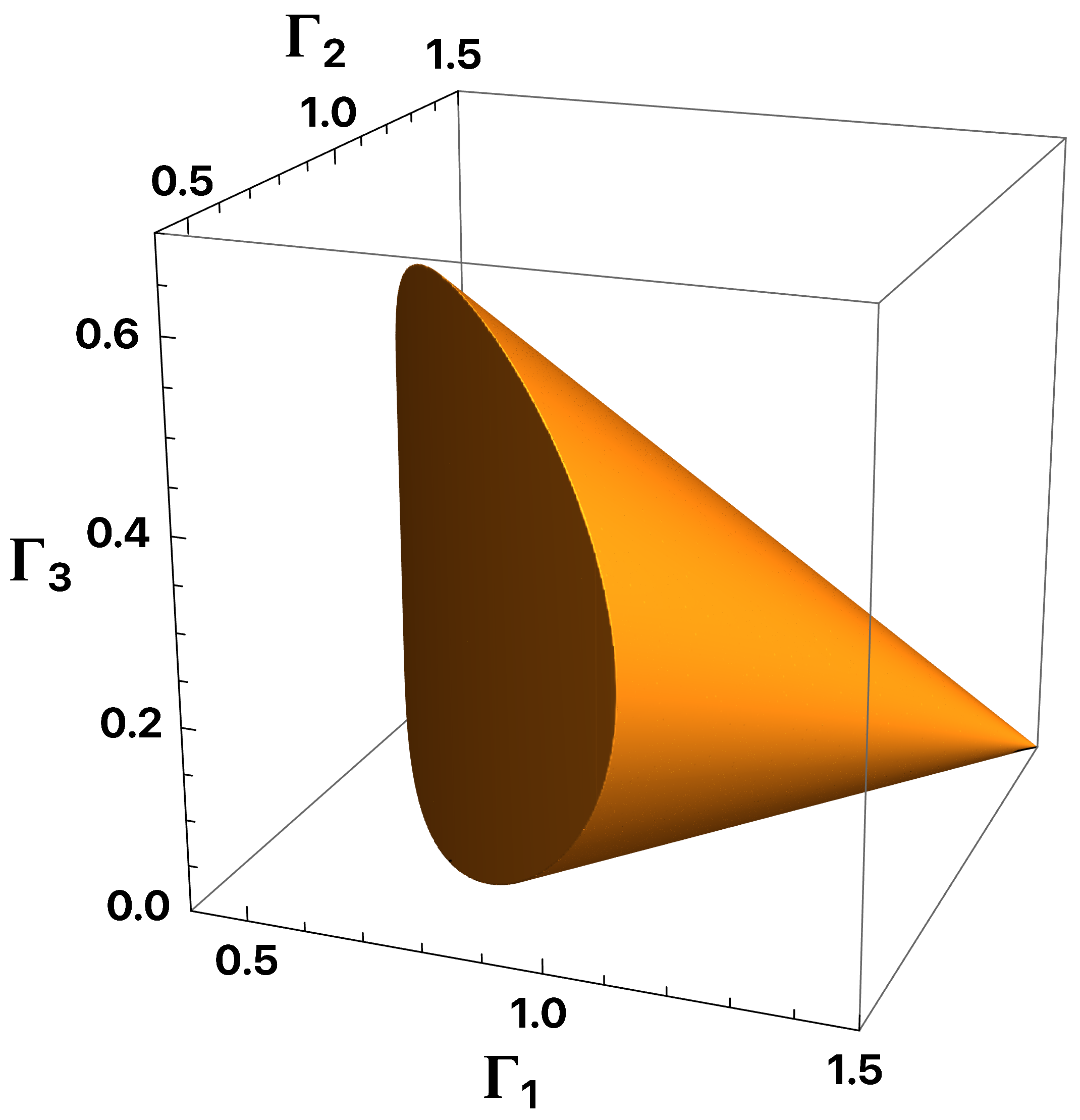}
    \caption{Region in the parameter space $(\Gamma_1,\,\Gamma_2,\,\Gamma_3)$ of the friction model that satisfies Eqs.~(\ref{cond1}), (\ref{cond2}) and 
    (\ref{cond3}) simultaneously, leading to positive semi-definite density matrices to first order in time. Here we set $n_\omega=g=\omega=\gamma=\hbar=1$.}
    \label{fig:fig2}
\end{figure}

For simplicity, we set $\Vec{\nu} = 0$. Consequently, the initial state~(\ref{F17}) is fully characterized by the positive, real and symmetric matrix
\begin{equation}
    \Vec{\Gamma}=\begin{pmatrix} \Gamma_1 & \Gamma_3 \\
                                 \Gamma_3 & \Gamma_2
    \end{pmatrix}\,.
\end{equation}
Positivity implies
\begin{equation}\label{cond1}
    \begin{cases}
        \Gamma_1+\Gamma_2>0\\
        \Gamma_1\Gamma_2>\Gamma_3^2
    \end{cases}\,,
\end{equation}
while the Heisenberg principle requires
\begin{equation}\label{cond2}
    \Gamma_1 \Gamma_2\geq \hbar/2\,.
\end{equation}
To linear order in time $t$, the positivity of the initial state~(\ref{F17}) is preserved if
\begin{equation}\label{condition_posit}
    \frac{\D\Vec{\Omega}_f(t)}{\D t}\bigg|_{t=0}=
    \begin{pmatrix} 2g\Gamma_3                                          & g\Gamma_2-\gamma\Gamma_3-\frac{\omega^2}{g}\Gamma_1 \\[0.2cm]
                    g\Gamma_2-\gamma\Gamma_3-\frac{\omega^2}{g}\Gamma_1 & \quad \frac{\omega\hbar\gamma}{g}(2n_\omega +1) - 2\gamma\Gamma_2-2\frac{\omega^2}{g}\Gamma_3
    \end{pmatrix}\,,
\end{equation}
is also a positive matrix, that implies 
\begin{equation}\label{cond3}
    \begin{cases}
        \frac{\omega\hbar\gamma}{g}(2n_\omega +1) - 2\gamma\Gamma_2-2\left(\frac{\omega^2}{g}-g\right)\Gamma_3>0\\[0.2cm]
        2g\Gamma_3\left[\frac{\omega\hbar\gamma}{g}(2n_\omega +1) - 2\gamma\Gamma_2-2\frac{\omega^2}{g}\Gamma_3\right]>\left[g\Gamma_2-\gamma\Gamma_3-\frac{\omega^2}{g}\Gamma_1\right]^2
    \end{cases}\,.
\end{equation}
Therefore, by combining conditions~(\ref{cond1},~\ref{cond2},~\ref{cond3}), we identify the region of the parameter space $(\Gamma_1,\,\Gamma_2,\,\Gamma_3)$ 
where the state remains positive semi-definite throughout the short-time evolution in the friction model. 
An example is provided in Fig.~\ref{fig:fig2}. It is worth mentioning that, while positivity is preserved in the short- and long-time regimes, it is not obvious that the same holds at intermediate times as well, despite it being the expected behaviour.

\section{Steady states}
\label{appendixE}

Eq.~(\ref{eq:W'}) implies that $\lim_{t\to\infty}\mathcal{W}'(t,\Vec{r})=\delta(\Vec{r})$ and thus (\ref{F:conv}) leads to 
\begin{equation}
    \lim_{t\to\infty}\mathcal{W}_{\alpha}(t,\Vec{r})=\int_{\mathbb{R}^2} \!\D \Vec{r'}\:  W_\alpha^{T}(\Vec{r}-\Vec{r'})\delta(\Vec{r'})=W_\alpha^{T}(\Vec{r})\,,
\end{equation}
where
\begin{equation}\label{eq:4.17b}
    W_\alpha^{T}(\Vec{r}):=\lim_{t\to\infty}f_\alpha(t,\Vec{r})=\frac{1}{2\pi\sqrt{\det\Vec{\Lambda}_T\,}\,}\exp{-\frac{1}{2}(\Vec{r}-\Vec{\mu}_\alpha)^T\Vec{\Lambda}_T^{-1}(\Vec{r}-\Vec{\mu}_\alpha)}\,.
\end{equation}
Therefore, the $2\times 2$ matrix $\Vec{\Lambda}_T$ and the vector $\Vec{\mu}_\alpha$ characterise the stationary state, 
with a one-to-one map~\cite{Polkovnikov2013} onto the Gibbs state $\hat{\rho}^T_\alpha$ of temperature $T$,
\begin{equation}
W_\alpha^{T}(\Vec{r}) ~~\longleftrightarrow~~ 
\hat{\rho}^T_\alpha = \frac{\e^{-\hat{\mathcal{H}}_{\alpha}/T}}{\tr(\e^{- \hat{\mathcal{H}}_{\alpha}/T})}\,,
\end{equation}
where 
\begin{equation}\label{eq:4.17ham}
\hat{\mathcal{H}}_f=\hat{H}\,,\qquad
\hat{\mathcal{H}}_c=\hat{H}+B\frac{\gamma^2}{\gamma^2+4\omega^2}\hat{x}-B\frac{2g\gamma}{\gamma^2+4\omega^2}\hat{p}   \,.
\end{equation}
To see this, recall that the equilibrium density matrix $\hat{\rho}^T_\alpha\propto \e^{-\hat{\mathcal{H}}_{\alpha}/T}$ 
corresponds to the Wigner function 
$W_\alpha^{T}(\Vec{r}) \propto \sum_{n=0}^{\infty} \frac{1}{n!} \bigl(-\mathcal{H}_{\alpha}(x,p)/T\bigr)^{\star n}$ 
with repeated Moyal products ($\star$) and $\mathcal{H}_{\alpha}(x,p)$ being the phase-space representation of $\hat{\mathcal{H}}_\alpha$. 
For the harmonic oscillator~(\ref{eq:H}) of the friction and cavity models, Eq.~\eqref{eq:4.17b} shows that 
$W_\alpha^{T}(\Vec{r})\propto \exp(- \mathcal{H}_{\alpha}(x,p)/T_{\rm eff})$, which takes indeed a harmonic oscillator form but with an effective temperature 
$T_{\rm eff}=\frac{\hbar\omega}{2} \coth\frac{\hbar\omega}{2T}$.
In the density matrix $\hat{\rho}_{\alpha}^{T}$, this gives back the bath temperature $T$~\cite{LandauLifschitz1979,PathriaBeale2011,Polkovnikov2013}. 

\newpage

\section*{References}
\bibliography{lib}

\end{document}